\documentclass[11pt]{article}%
\usepackage{amsmath}
\usepackage{amssymb}
\usepackage{amsfonts}

\usepackage{cite}
\usepackage{graphicx}
\usepackage{float}
\usepackage{longtable}
\usepackage[utf8]{inputenc}
\usepackage{hyperref}

\usepackage{algorithmicx}
\usepackage[ruled,vlined]{algorithm2e}
\SetAlgoCaptionSeparator{ }

\usepackage{fullpage}
\usepackage{authblk}

\newcommand{\fref}[1]{Fig.~\ref{#1}}

\newcommand{\tref}[1]{Table~\ref{#1}}
\newcommand{\sref}[1]{Section~\ref{#1}}

\newenvironment{algo}[1][!htbp]
  {
   \begin{algorithm}[#1]%
  }{\end{algorithm}}

\providecommand{\U}[1]{\protect\rule{.1in}{.1in}}
\newtheorem{theorem}{Theorem}

\newtheorem{lemma}{Lemma}

\newenvironment{proof}[1][Proof]{\textbf{#1.} }{\ \rule{0.5em}{0.5em}}

\begin{document}

\title{Multiple Access Multicarrier Continuous-Variable Quantum Key Distribution}
\author[1,2,3]{Laszlo Gyongyosi\footnote{Email: \href{mailto:l.gyongyosi@soton.ac.uk}{l.gyongyosi@soton.ac.uk}. Parts of this work were presented in conference proceedings [4].}}
\author[2]{Sandor Imre}
\affil[1]{School of Electronics and Computer Science, University of Southampton, Southampton, SO17 1BJ, UK}
\affil[2]{Department of Networked Systems and Services, Budapest University of Technology and Economics, Budapest, H-1117 Hungary}
\affil[3]{MTA-BME Information Systems Research Group, Hungarian Academy of Sciences, Budapest, H-1051 Hungary}
\date{}

\maketitle
\begin{abstract}
One of the most important practical realizations of the fundamentals of quantum mechanics is continuous-variable quantum key distribution (CVQKD). Here we propose the adaptive multicarrier quadrature division–multiuser quadrature allocation (AMQD-MQA) multiple access technique for continuous-variable quantum key distribution. The MQA scheme is based on the AMQD modulation, which granulates the inputs of the users into Gaussian subcarrier continuous-variables (CVs). In an AMQD-MQA multiple access scenario, the simultaneous reliable transmission of the users is handled by the dynamic allocation of the Gaussian subcarrier CVs. We propose two different settings of AMQD-MQA for multiple input–multiple output communication. We introduce a rate-selection strategy that tunes the modulation variances and allocates adaptively the quadratures of the users over the sub-channels. We also prove the rate formulas if only partial channel side information is available for the users of the sub-channel conditions. We show a technique for the compensation of a nonideal Gaussian input modulation, which allows the users to overwhelm the modulation imperfections to reach optimal capacity-achieving communication over the Gaussian sub-channels. We investigate the diversity amplification of the sub-channel transmittance coefficients and reveal that a strong diversity can be exploited by opportunistic Gaussian modulation.
\end{abstract}

\section{Introduction}
The continuous-variable quantum key distribution (CVQKD) protocols allow the parties to realize unconditionally secure communication over the standard, currently established telecommunication networks \cite{ref1,ref2,ref3,ref4,ref5,ref6,ref7,ref8,ref9,ref10,ref11,ref12,ref13,ref14,ref15,ref16}. The CVQKD schemes have several benefits over the discrete variable (DV) quantum key distribution (QKD) protocols; most importantly, they do not require single-photon encoding and decoding, which allows its practical implementation by standard, currently available technologies and devices \cite{ref1,ref2,ref3,ref4,ref5,ref6,ref7,ref8,ref9,ref10,ref11,ref12,ref13,ref14,ref36}. The CVQKD protocols use continuous-variables (CVs) for the information transmission, practically Gaussian-modulated quantum coherent states in the phase space. The Gaussian modulation is particularly convenient in an experimental scenario, which allows the parties to communicate over a quantum channel by Gaussian random distributed position and momentum quadratures. The communication link that connects the parties can be a standard optical fiber or an optical wireless channel \cite{ref24,ref25,ref26,ref27,ref28,ref29,ref30,ref31,ref32,ref33,ref34,ref35,ref36,ref37,ref38,ref39,ref40,ref41,ref42,ref43,ref44,ref45,ref46}. The quantum link is attacked by an eavesdropper, which adds a white Gaussian noise into the transmission. (In particular, the optimal attack against CVQKD is a Gaussian attack \cite{ref12,ref13}.) As follows, the physical quantum link can be provably modeled as a Gaussian quantum channel. The currently developed CVQKD protocols are based on single-carrier Gaussian CV states, where the sender, Alice, modulates separately the Gaussian coherent states, which are then transmitted through a noisy quantum channel. The noisy Gaussian CVs are then decoded by the receiver, Bob. The single-carrier modulation does not perform such advanced techniques within CVQKD as it is already available in a traditional telecommunication scenario. As a corollary, several important communication techniques cannot be implemented within the framework of the CVQKD protocols. To eliminate these drawbacks, the adaptive multicarrier quadrature division (AMQD) modulation has been recently introduced \cite{ref4,ref23,ref24,ref25}, which allows the parties to significantly extend the possibilities of single-carrier CVQKD protocols. 

The AMQD is based on the use of the Gaussian subcarrier CVs and continuous unitary operations and offers several benefits over the single-carrier modulation. It provides higher noise resistance, higher tolerable loss, improved rates, and transmission distances for the parties. The question is now straightforward. The multiple access scheme, which is introduced with this in mind, is called adaptive multicarrier quadrature division--multiuser quadrature allocation (AMQD-MQA). The AMQD-MQA scheme exploits and extends the benefits of AMQD modulation into a multiple access scenario. The AMQD-MQA allows the realization of multiple input--multiple output transmission within CVQKD, making it possible for users to have a simultaneous reliable communication over the physical Gaussian quantum channel by the dynamic allocation of the Gaussian subcarrier CVs. In particular, the AMQD-MQA transmits the information of the users via modulated Gaussian subcarrier CVs, where each subcarrier is allocated with a constant modulation variance. It is provably the optimal solution in the low-SNR (signal to noise ratio) regimes, which is specifically the case in experimental long-distance CVQKD. In the AMQD-MQA, the rate of the users is allocated by the MQA technique, which integrates the utilization of continuous unitary operations and the sophisticated management of the Gaussian-modulated subcarrier CVs. The subcarrier CVs divide the physical Gaussian channel into Gaussian sub-channels, each dedicated for the transmission of a given Gaussian subcarrier with an independent noise variance. The inputs of the selected independent transmit users are conveyed by the Gaussian subcarrier CV states, which are received by the independent parties using an inverse continuous unitary. The noise acts on the position and momentum quadratures of the Gaussian subcarrier CVs. The AMQD-MQA is equipped with all the benefits of AMQD, such as improved tolerable loss and excess noise, higher transmission distances, and optimized key rates. It extends the possibilities of AMQD for a multiuser scenario, which allows all users to simultaneously achieve the benefits provided by the AMQD framework similar to the well-known orthogonal frequency-division multiplexing multiple access (OFDMA) of traditional networking \cite{ref9,ref15,ref17,ref18,ref19,ref20,ref21,ref22}. As an important corollary, the AMQD-MQA overwhelms the problems of single-carrier protocols to reach a much more efficient and significantly optimized multiple access transmission compared with a single-carrier multiuser scheme.

The AMQD-MQA multiple access uses Gaussian modulation, which allows easy implementation in a practical scenario by standard devices. We prove the achievable user rates in AMQD-MQA multiple access and the optimality of the scheme. The AMQD modulation does not require the exact tracking of the sub-channel conditions, allowing the use of constant modulation variance throughout the modulation of the subcarriers. As a fine corollary, the AMQD-MQA provides capacity-achieving communication only if partial channel side information is available for the parties, which is particularly convenient in an experiential long-distance scenario. The Gaussian modulation is a well-applicable solution in practice; however, an ideal, perfect Gaussian modulation can only be approximated, which causes deviations in the input distribution. 

Here we show a technique for the compensation of nonideal Gaussian modulation, which can be used to further improve the performance of the AMQD-MQA. The Gaussian sub-channels that transmit the subcarriers are characterized by a given transmittance coefficient, which models the position and momentum quadrature transmission process over the Gaussian channel. Finally, we introduce the diversity amplification technique for the Gaussian sub-channels. The diversity of the transmittance coefficients of the sub-channels can be significantly improved by opportunistic Gaussian modulation. This type of modulation randomizes the distribution of the coefficients that results in an extended distribution range and improved SNRs for the users.

This paper is organized as follows. In \sref{sec2}, the preliminaries are summarized. \sref{sec3} discusses the details of the AMQD-MQA mechanism. In \sref{sec4}, we show a technique for the compensation of nonideal Gaussian-modulated inputs. \sref{sec5} investigates the diversity amplification by opportunistic Gaussian modulation within AMQD-MQA. Finally, \sref{sec6} concludes the results.

\section{AMQD-MQA}
\label{sec2}
\subsection{Preliminaries}
First we provide the system initialization of AMQD-MQA, following the parameterization of AMQD modulation framework \cite{ref4}. The variables of the AMQD-MQA protocol are summarized as follows. The input of $k$-th user $U_{k} $ is a Gaussian CV state ${\left| \psi  \right\rangle} _{k} \in {\rm {\mathcal S}}$, where ${\rm {\mathcal S}}$ stands for the phase space. The noise of the Gaussian quantum channel ${\rm {\mathcal N}}$ is denoted by $\Delta $, which acts independently on the $x$ position, and $p$ momentum quadratures. These Gaussian CV states, along with the noise of the quantum channel, can be modeled as Gaussian random continuous-variables. A single-carrier Gaussian modulated CV state ${\left| \psi  \right\rangle} $ in the phase space ${\rm {\mathcal S}}$ is modeled as a zero-mean, circular symmetric complex Gaussian random variable $z\in {\rm {\mathcal C}{\mathcal{N}}}\left(0,\sigma _{z}^{2} \right),$ with i.i.d. zero mean, Gaussian random quadrature components $x,p\in {\rm {\mathcal{N}}}\left(0,\sigma _{\omega _{0} }^{2} \right)$, where $\sigma _{\omega _{0} }^{2} $ is the modulation variance. The variance of $z$ is 
\begin{equation} \label{1)} 
\sigma _{z}^{2} ={\rm {\mathbb{E}}}\left[\left|z\right|^{2} \right]=2\sigma _{\omega _{0} }^{2} .                                            
\end{equation} 
The $\Delta \in {\rm {\mathcal C}{\mathcal{N}}}\left(0,\sigma _{\Delta }^{2} \right)$ noise variable of the Gaussian channel ${\rm {\mathcal N}}$ with i.i.d. zero-mean, Gaussian random components on the position and momentum quadratures $\Delta _{x} ,\Delta _{p} \in {\rm {\mathcal{N}}}\left(0,\sigma _{{\rm {\mathcal N}}}^{2} \right)$ has the variance of 
\begin{equation} \label{2)} 
\sigma _{\Delta }^{2} ={\rm {\mathbb{E}}}\left[\left|\Delta \right|^{2} \right]=2\sigma _{{\rm {\mathcal N}}}^{2} .                                          
\end{equation} 
A ${| \phi \rangle} \in {\rm {\mathcal S}}$ Gaussian subcarrier CV state is also modeled by a zero-mean, circular symmetric Gaussian random variable $d\in {\rm {\mathcal C}{\mathcal{N}}}\left(0,\sigma _{d}^{2} \right),$ with i.i.d. zero mean, Gaussian random quadrature components $x_{d}, p_{d} \in {\rm {\mathcal{N}}}\left(0,\sigma _{\omega }^{2} \right)$, where $\sigma _{\omega }^{2} $ is the modulation variance of the Gaussian subcarrier CV state. The Gaussian subcarrier random variable $d$ has variance
\begin{equation} \label{3)} 
\sigma _{d}^{2} ={\rm {\mathbb{E}}}\left[\left|d\right|^{2} \right]=2\sigma _{\omega }^{2} .                                             
\end{equation} 
Let assume that $K$ independent users are in the CVQKD system. 

The subset of allocated users is denoted by ${\rm {\mathcal A}}\subseteq K$. Only the allocated users can transmit information in a given (particularly $j$-th) AMQD block. An AMQD block consist of $l$ Gaussian subcarrier CVs (assuming an optimal Gaussian collective attack \cite{ref12,ref13}, only $l$ sub-channels have high quality from the n, these are referred as good sub-channels throughout, for details see \cite{ref4}). 

Following the formalism of AMQD \cite{ref4}, the variables of an allocated user $U_{k} ,{\rm \; }k=1,\ldots ,\left|{\rm {\mathcal A}}\right|$, where $\left|{\rm {\mathcal A}}\right|$ is the cardinality of the subset ${\rm {\mathcal A}}$, are as follows. The $i$-th Gaussian modulated input coherent state of $U_{k} $ is referred as ${\left| \varphi _{k,i}  \right\rangle} ={\left| x_{k,i} +{\rm i}p_{k,i}  \right\rangle} $, where $x_{k,i} \in {\rm {\mathcal{N}}}\left(0,\sigma _{\omega _{0,k} }^{2} \right)$, $p_{k,i} \in {\rm {\mathcal{N}}}\left(0,\sigma _{\omega _{0,k} }^{2} \right)$ are the position and momentum quadratures with variance $\sigma _{\omega _{0,k} }^{2} $, respectively. This CV state can be rewritten as a zero-mean, circular symmetric complex Gaussian random variable $z_{k,i} \in {\rm {\mathcal C}{\mathcal{N}}}\left(0,\sigma _{\omega _{z_{k,i} } }^{2} \right)$, $\sigma _{\omega _{z_{k,i} } }^{2} =\left[{\rm {\mathbb{E}}}\left|z_{k,i} \right|^{2} \right]$, as
\begin{equation} \label{4)} 
z_{k,i} =x_{k,i} +{\rm i}p_{k,i} ,                                             
\end{equation} 
thus 
\begin{equation} \label{5)} 
{\left| \varphi _{k,i}  \right\rangle} ={\left| z_{k,i}  \right\rangle} .                                               
\end{equation} 
The variable $e^{{\rm i}\varphi _{i} } z_{k,i} $ has the same distribution of $z_{k,i} $ for any $\varphi _{i} $, i.e., ${\rm {\mathbb{E}}}\left[z_{k,i} \right]={\rm {\mathbb{E}}}\left[e^{{\rm i}\varphi _{i} } z_{k,i} \right]={\rm {\mathbb{E}}}e^{{\rm i}\varphi _{i} } \left[z_{k,i} \right]$ and $\sigma _{z_{k,i} }^{2} ={\rm {\mathbb{E}}}\left[\left|z_{k,i} \right|^{2} \right]$. The density of $z_{k,i} $ is
\begin{equation} \label{ZEqnNum239770} 
f\left(z_{k,i} \right)={\textstyle\frac{1}{2\pi \sigma _{\omega _{0,k} }^{2} }} e^{{\textstyle\frac{-\left(\left|z_{k,i} \right|^{2} \right)}{2\sigma _{\omega _{0,k} }^{2} }} } =f\left(x_{k,i} ,p_{k,i} \right)={\textstyle\frac{1}{2\pi \sigma _{\omega _{0,k} }^{2} }} e^{{\textstyle\frac{-\left(x_{k,i}^{2} +p_{k,i}^{2} \right)}{2\sigma _{\omega _{0,k} }^{2} }} } , 
\end{equation} 
where $\left|z_{k,i} \right|=\sqrt{x_{k,i}^{2} +p_{k,i}^{2} } $ is the magnitude, which is a Rayleigh random variable with density
\begin{equation} \label{7)} 
f\left(\left|z_{k,i} \right|\right)={\textstyle\frac{\left|z_{k,i} \right|}{\sigma _{\omega _{z_{k,i} } }^{2} }} e^{{\textstyle\frac{-\left|z_{k,i} \right|^{2} }{2\sigma _{\omega _{z_{k,i} } }^{2} }} } ,\left|z_{k,i} \right|\ge 0,                                     
\end{equation} 
while the $\left|z_{k,i} \right|^{2} =x_{k,i}^{2} +p_{k,i}^{2} $ squared magnitude is exponentially distributed with density
\begin{equation} \label{8)} 
f\left(\left|z_{k,i} \right|^{2} \right)={\textstyle\frac{1}{\sigma _{\omega _{z_{k,i} } }^{2} }} e^{{\textstyle\frac{-\left|z_{k,i} \right|^{2} }{\sigma _{\omega _{z_{k,i} } }^{2} }} } ,\left|z_{k,i} \right|^{2} \ge 0.                                  
\end{equation} 
The $i$-th Gaussian subcarrier CV of user $U_{k} $ is defined as 
\begin{equation} \label{ZEqnNum152136} 
{\left| \phi _{i}  \right\rangle} ={\left| {\rm IFFT}\left(z_{k,i} \right) \right\rangle} ={\left| F^{-1} \left(z_{k,i} \right) \right\rangle} ={\left| d_{i}  \right\rangle} ,                              
\end{equation} 
where IFFT stands for the inverse fast Fourier transform, and subcarrier continuous-variable ${\left| \phi _{i}  \right\rangle} $ in Equation \eqref{ZEqnNum152136} is also a zero-mean, circular symmetric complex Gaussian random variable $d_{i} \in {\rm {\mathcal C}{\mathcal{N}}}\left(0,\sigma _{d_{i} }^{2} \right)$, $\sigma _{d_{i} }^{2} ={\rm {\mathbb{E}}}\left[\left|d_{i} \right|^{2} \right]$. The quadrature components of the modulated Gaussian subcarrier CVs are referred by $x_{d} ,p_{d} \in {\rm {\mathcal{N}}}\left(0,\sigma _{\omega }^{2} \right)$, where $\sigma _{\omega }^{2} $ is the constant modulation variance of AMQD that is used in the transmission phase (Note: the constant modulation variance is provably the optimal solution in the low-SNR regimes, because the performance is very close to the exact allocation \cite{ref4, ref20,ref21,ref22}.).

Assuming $K$ independent users in the AMQD-MQA, who transmit the single-carrier CVs to an encoder ${\rm {\mathcal E}}$, \eqref{ZEqnNum152136} modifies as follows (see \fref{fig2})
\begin{equation} \label{10)} 
{\left| \phi _{i}  \right\rangle} ={\left| {\text{CVQFT}}^{\dag } \left(z_{k,i} \right) \right\rangle} ={\left| F^{-1} \left(z_{k,i} \right) \right\rangle} ={\left| d_{i}  \right\rangle} ,                        
\end{equation} 
where ${\text{CVQFT}}^{\dag } $ refers to the continuous-variable inverse quantum Fourier transform (QFT).

The inverse of \eqref{ZEqnNum152136} results the single-carrier CV from the subcarrier CV as follows:
\begin{equation} \label{11)} 
\left| {{\varphi }_{k,i}} \right\rangle =\text{CVQFT}\left( \left| {{\phi }_{i}} \right\rangle  \right)=F\left( \left| {{d}_{i}} \right\rangle  \right)=\left| F\left( {{F}^{-1}}\left( {{z}_{k,i}} \right) \right) \right\rangle =\left| {{z}_{k,i}} \right\rangle ,             
\end{equation} 
where $\text{CVQFT}$ is the continuous-variable QFT operation. 

Let ${{\mathbf{z}}_{k}}$ be an $d$-dimensional, zero-mean, circular symmetric complex random Gaussian vector of $U_{k} $,
\begin{equation} \label{ZEqnNum732901} 
{{\mathbf{z}}_{k}}={{\mathbf{x}}_{k}}+i{{\mathbf{p}}_{k}}={{\left( {{z}_{k,1}},\ldots ,{{z}_{k,d}} \right)}^{T}}\in \mathcal{C}\mathcal{N}\left( 0,{{\mathbf{K}}_{{{\mathbf{z}}_{k}}}} \right) ,                  
\end{equation} 
where ${{\mathbf{K}}_{{{\mathbf{z}}_{k}}}} $ is the $d\times d$ Hermitian covariance matrix of ${{\mathbf{z}}_{k}}$, ${{\mathbf{K}}_{{{\mathbf{z}}_{k}}}}=\mathbb{E}\left[ {{\mathbf{z}}_{k}}\mathbf{z}_{k}^{\dagger } \right]$, and $z_{k}^{\dag } $ stands for the adjoint of ${{\mathbf{z}}_{k}}$. Each $z_{k,i} $ variable is a zero-mean, circular symmetric complex Gaussian random variable $z_{k,i} \in {\rm {\mathcal C}{\mathcal{N}}}\left(0,\sigma _{\omega _{z_{k,i} } }^{2} \right)$, $z_{k,i} =x_{k,i} +{\rm i}p_{k,i} $. The real and imaginary variables (i.e., the position and momentum quadratures) formulate $d$-dimensional real Gaussian random vectors, ${{\mathbf{x}}_{k}}={{\left( {{x}_{k,1}},\ldots ,{{x}_{k,d}} \right)}^{T}}$ and ${{\mathbf{p}}_{k}}={{\left( {{p}_{k,1}},\ldots ,{{p}_{k,d}} \right)}^{T}}$, with zero-mean Gaussian random variables 
\begin{equation} \label{13)} 
x_{k,i} ={\textstyle\frac{1}{\sigma _{\omega _{0,k} } \sqrt{2\pi } }} e^{{\textstyle\frac{-x_{k,i} ^{2} }{2\sigma _{\omega _{0,k} }^{2} }} } , p_{k,i} ={\textstyle\frac{1}{\sigma _{\omega _{0,k} } \sqrt{2\pi } }} e^{{\textstyle\frac{-p_{k,i} ^{2} }{2\sigma _{\omega _{0,k} }^{2} }} } ,                            
\end{equation} 
where $\sigma _{\omega _{0,k} }^{2} $ is the stands for single-carrier modulation variance (precisely, the variance of the real and imaginary components of $z_{k,i} $). For vector ${{\mathbf{z}}_{k}}$, 
\begin{equation} \label{14)} 
\mathbb{E}\left[ {{\mathbf{z}}_{k}} \right]=\mathbb{E}\left[ {{e}^{i\gamma }}{{\mathbf{z}}_{k}} \right]=\mathbb{E}{{e}^{i\gamma }}\left[ {{\mathbf{z}}_{k}} \right]
\end{equation} 
holds, and
\begin{equation} \label{15)} 
\mathbb{E}\left[ {{\mathbf{z}}_{k}}\mathbf{z}_{k}^{T} \right]=\mathbb{E}\left[ {{e}^{i\gamma }}{{\mathbf{z}}_{k}}{{\left( {{e}^{i\gamma }}{{\mathbf{z}}_{k}} \right)}^{T}} \right]=\mathbb{E}{{e}^{i2\gamma }}\left[ {{\mathbf{z}}_{k}}\mathbf{z}_{k}^{T} \right]
\end{equation} 
for any $\gamma \in \left[0,2\pi \right]$. The density of ${{\mathbf{z}}_{k}}$ is as follows (if ${{\mathbf{K}}_{{{\mathbf{z}}_{k}}}}$ is invertible):
\begin{equation} \label{16)} 
f\left( {{\mathbf{z}}_{k}} \right)=\tfrac{1}{{{\pi }^{d}}\det {{\mathbf{K}}_{{{\mathbf{z}}_{k}}}}}{{e}^{-\mathbf{z}_{k}^{\dagger }\mathbf{K}_{{{\mathbf{z}}_{k}}}^{-1}{{\mathbf{z}}_{k}}}}.
\end{equation} 
A $d$-dimensional Gaussian random vector is expressed as ${{\mathbf{x}}_{k}}=\mathbf{As}$, where $\mathbf{A}$ is an (invertible) linear transform from ${{\mathbb{R}}^{d}}$ to ${{\mathbb{R}}^{d}}$, and $\mathbf{s}$ is a $d$-dimensional standard Gaussian random vector ${\rm {\mathcal{N}}}\left(0,1\right)_{d} $. This vector is characterized by its covariance matrix ${{\mathbf{K}}_{{{\mathbf{x}}_{k}}}}=\mathbb{E}\left[ {{\mathbf{x}}_{k}}\mathbf{x}_{k}^{T} \right]=\mathbf{A}{{\mathbf{A}}^{T}}$, as
\begin{equation} \label{17)} 
{{\mathbf{x}}_{k}}=\tfrac{1}{{{\left( \sqrt{2\pi } \right)}^{d}}\sqrt{\det \left( \mathbf{A}{{\mathbf{A}}^{T}} \right)}}{{e}^{-\tfrac{\mathbf{x}_{k}^{T}{{\mathbf{x}}_{k}}}{2\left( \mathbf{A}{{\mathbf{A}}^{T}} \right)}}}.
\end{equation} 
The Fourier transformation $F\left(\cdot \right)$ of an $l$-dimensional Gaussian random vector $\mathbf{v}={{\left( {{v}_{1}},\ldots ,{{v}_{l}} \right)}^{T}}$ results in the $d$-dimensional Gaussian random vector $\mathbf{m}={{\left( {{m}_{1}},\ldots ,{{m}_{d}} \right)}^{T}}$:
\begin{equation} \label{18)} 
\mathbf{m}=F\left( \mathbf{v} \right)={{e}^{\tfrac{-{{\mathbf{m}}^{T}}\mathbf{A}{{\mathbf{A}}^{T}}\mathbf{m}}{2}}}={{e}^{\tfrac{-\sigma _{{{\omega }_{0}}}^{2}\left( m_{1}^{2}+\ldots +m_{d}^{2} \right)}{2}}}.
\end{equation} 
In the first step of AMQD, Alice applies the inverse FFT operation to vector ${{\mathbf{z}}_{k}}$ (see Equation \eqref{ZEqnNum732901}), which outputs an $l$-dimensional zero-mean, circular symmetric complex Gaussian random vector $\mathbf{d}$, $\mathbf{d}\in \mathcal{C}\mathcal{N}\left( 0,{{\mathbf{K}}_{\mathbf{d}}} \right)$, $\mathbf{d}={{\left( {{d}_{1}},\ldots ,{{d}_{l}} \right)}^{T}}$, as
\begin{equation} \label{ZEqnNum999736} 
\mathbf{d}={{F}^{-1}}\left( {{\mathbf{z}}_{k}} \right)={{e}^{\tfrac{{{\mathbf{d}}^{T}}\mathbf{A}{{\mathbf{A}}^{T}}\mathbf{d}}{2}}}={{e}^{\tfrac{\sigma _{{{\omega }_{0}}}^{2}\left( d_{1}^{2}+\ldots +d_{l}^{2} \right)}{2}}},
\end{equation} 
where $d_{i} =x_{d_{i} } +{\rm i}p_{d_{i} } $, $d_{i} \in {\rm {\mathcal C}{\mathcal{N}}}\left(0,\sigma _{d_{i} }^{2} \right)$, and the position and momentum quadratures of ${\left| \phi _{i}  \right\rangle} $ are i.i.d. Gaussian random variables
\begin{equation} \label{20)} 
x_{d_{i} } \in {\rm {\mathcal{N}}}\left(0,\sigma _{F}^{2} \right), p_{d_{i} } \in {\rm {\mathcal{N}}}\left(0,\sigma _{F}^{2} \right),                                   
\end{equation} 
where ${{\mathbf{K}}_{\mathbf{d}}}=\mathbb{E}\left[ \mathbf{d}{{\mathbf{d}}^{\dagger }} \right]$, $\mathbb{E}\left[ \mathbf{d} \right]=\mathbb{E}\left[ {{e}^{i\gamma }}\mathbf{d} \right]=\mathbb{E}{{e}^{i\gamma }}\left[ \mathbf{d} \right]$, and $\mathbb{E}\left[ \mathbf{d}{{\mathbf{d}}^{T}} \right]=\mathbb{E}\left[ {{e}^{i\gamma }}\mathbf{d}{{\left( {{e}^{i\gamma }}\mathbf{d} \right)}^{T}} \right]=\mathbb{E}{{e}^{i2\gamma }}\left[ \mathbf{d}{{\mathbf{d}}^{T}} \right]$ for any $\gamma \in \left[0,2\pi \right]$. The coherent Gaussian subcarrier CV is as follows:
\begin{equation} \label{21)} 
{\left| \phi _{i}  \right\rangle} ={\left| d_{i}  \right\rangle} ={\left| F^{-1} \left(z_{k} \right) \right\rangle} .                                     
\end{equation} 
The result of Equation \eqref{ZEqnNum999736} defines $l$ independent ${\rm {\mathcal N}}_{i} $ Gaussian sub-channels, each with noise variance $\sigma _{{\rm {\mathcal N}}_{i} }^{2} $, one for each subcarrier coherent state ${\left| \phi _{i}  \right\rangle} $. After the CV subcarriers are transmitted through the noisy quantum channel, Bob applies the CVQFT, which results him the noisy version ${\left| \varphi '_{k}  \right\rangle} ={\left| z'_{k}  \right\rangle} $ of the input $z_{k} $ of $U_{k} $. 

The $m$-th element of $d$-dimensional zero-mean, circular symmetric complex Gaussian random output vector ${{\mathbf{y}}_{k}}\in \mathcal{C}\mathcal{N}\left( 0,\mathbb{E}\left[ {{\mathbf{y}}_{k}}\mathbf{y}_{k}^{\dagger } \right] \right)$ of $U_{k} $, is as follows:
\begin{equation} \label{ZEqnNum449265} 
\begin{split}
   {{y}_{k,m}}&=F\left( \mathbf{T}\left( \mathcal{N} \right) \right){{z}_{k,m}}+F\left( \Delta  \right) \\ 
 & =F\left( \mathbf{T}\left( \mathcal{N} \right) \right)F\left( {{F}^{-1}}\left( {{z}_{k,m}} \right) \right)+F\left( \Delta  \right)\text{ } \\ 
 & =\sum\nolimits_{l}{F\left( {{T}_{i}}\left( {{\mathcal{N}}_{i}} \right) \right)}F\left( {{d}_{i}} \right)+F\left( {{\Delta }_{i}} \right),  
\end{split}
\end{equation} 
where $F\left( \mathbf{T}\left( \mathcal{N} \right) \right)$ is the Fourier transform of the $l$-dimensional complex channel transmission vector 
\begin{equation} \label{ZEqnNum348185} 
\mathbf{T}\left( \mathcal{N} \right)={{\left[ {{T}_{1}}\left( {{\mathcal{N}}_{1}} \right)\ldots ,{{T}_{l}}\left( {{\mathcal{N}}_{l}} \right) \right]}^{T}}\in {{\mathcal{C}}^{l}},
\end{equation} 
where
\begin{equation} \label{ZEqnNum541779} 
T_{i} \left({\rm {\mathcal N}}_{i} \right)=\text{Re}\left(T_{i} \left({\rm {\mathcal N}}_{i} \right)\right)+{\rm i}\text{Im}\left(T_{i} \left({\rm {\mathcal N}}_{i} \right)\right)\in {\rm {\mathcal C}},                        
\end{equation} 
is a complex variable, called transmittance coefficient, which quantifies the position and momentum quadrature transmission (i.e., gain) of the $i$-th Gaussian sub-channel ${\rm {\mathcal N}}_{i} $, in the phase space ${\rm {\mathcal S}}$, with (normalized) real and imaginary parts $0\le \text{Re}T_{i} \left({\rm {\mathcal N}}_{i} \right)\le {1 \mathord{\left/{\vphantom{1 \sqrt{2} }}\right.\kern-\nulldelimiterspace} \sqrt{2} } ,$ $0\le \text{Im}T_{i} \left({\rm {\mathcal N}}_{i} \right)\le {1 \mathord{\left/{\vphantom{1 \sqrt{2} }}\right.\kern-\nulldelimiterspace} \sqrt{2} } $. The $T_{i} \left({\rm {\mathcal N}}_{i} \right)$ variable has a magnitude of $\left|T_{i} \left({\rm {\mathcal N}}_{i} \right)\right|=\sqrt{\text{Re}T_{i} \left({\rm {\mathcal N}}_{i} \right)^{2} +\text{Im}T_{i} \left({\rm {\mathcal N}}_{i} \right)^{2} } \in {\rm {\mathbb R}}$, where $\text{Re}T_{i} \left({\rm {\mathcal N}}_{i} \right)=\text{Im}T_{i} \left({\rm {\mathcal N}}_{i} \right)$, by our convention.  

For the $l$ sub-channels, the $F\left(\Delta \right)$ complex vector is evaluated as
\begin{equation} \label{ZEqnNum334769} 
F\left(\Delta \right)=e^{{\textstyle\frac{-F\left(\Delta \right)^{T} K_{F\left(\Delta \right)} F\left(\Delta \right)}{2}} } =e^{{\textstyle\frac{-\left(F\left(\Delta _{1} \right)^{2} \sigma _{_{{\rm {\mathcal N}}_{1} } }^{2} +\ldots +F\left(\Delta _{l} \right)^{2} \sigma _{_{{\rm {\mathcal N}}_{l} } }^{2} \right)}{2}} } , 
\end{equation} 
which is the Fourier transform of the $l$-dimensional zero-mean, circular symmetric complex Gaussian noise vector $\Delta \in {\rm {\mathcal C}{\mathcal{N}}}\left(0,\sigma _{\Delta }^{2} \right)_{d} $, $\Delta ={{\left( {{\Delta }_{1}},\ldots ,{{\Delta }_{l}} \right)}^{T}}\in \mathcal{C}\mathcal{N}\left( 0,{{\mathbf{K}}_{\Delta }} \right)$, where ${{\mathbf{K}}_{\Delta }}=\mathbb{E}\left[ \Delta {{\Delta }^{\dagger }} \right]$, with independent, zero-mean Gaussian random components $\Delta _{x_{i} } \in {\rm {\mathcal{N}}}\left(0,\sigma _{{\rm {\mathcal N}}_{i} }^{2} \right)$, $\Delta _{p_{i} } \in {\rm {\mathcal{N}}}\left(0,\sigma _{{\rm {\mathcal N}}_{i} }^{2} \right)$ with variance $\sigma _{{\rm {\mathcal N}}_{i} }^{2} $ for each $\Delta _{i} $. These identify the Gaussian noise of sub-channel ${\rm {\mathcal N}}_{i} $ on the quadrature components in the phase space ${\rm {\mathcal S}}$. The CVQFT-transformed noise vector can be rewritten as $F\left(\Delta \right)=\left(F\left(\Delta _{1} \right),\ldots ,F\left(\Delta _{l} \right)\right)^{T} $, $F\left(\Delta _{x_{i} } \right)\in {\rm {\mathcal{N}}}\left(0,\sigma _{F\left({\rm {\mathcal N}}_{i} \right)}^{2} \right)$, $F\left(\Delta _{p_{i} } \right)\in {\rm {\mathcal{N}}}\left(0,\sigma _{F\left({\rm {\mathcal N}}_{i} \right)}^{2} \right)$ for each $F\left(\Delta _{i} \right)$, which defines a $d$-dimensional zero-mean, circular symmetric complex Gaussian random vector $F\left( \Delta  \right)\in \mathcal{C}\mathcal{N}\left( 0,{{\mathbf{K}}_{F\left( \Delta  \right)}} \right)$ with a covariance matrix
\begin{equation} \label{ZEqnNum641952} 
{{\mathbf{K}}_{F\left( \Delta  \right)}}=\mathbb{E}\left[ F\left( \Delta  \right)F{{\left( \Delta  \right)}^{\dagger }} \right].
\end{equation} 
An AMQD block is formulated from $l$ Gaussian subcarrier continuous-variables, as follows: 
\begin{equation} \label{27)} 
\mathbf{y}\left[ j \right]=F\left( \mathbf{T}\left( \mathcal{N} \right) \right)F\left( \mathbf{d} \right)\left[ j \right]+F\left( \Delta  \right)\left[ j \right],
\end{equation} 
where $j$ is the index of the AMQD block, $F\left( \mathbf{d} \right)=F\left( {{F}^{-1}}\left( \mathbf{z} \right) \right)$, for ${{F}^{-1}}\left( \mathbf{z} \right)$ see \eqref{ZEqnNum999736}, while
\begin{equation} \label{ZEqnNum736005} 
\begin{split}
\mathbf{y}\left[ j \right]&={{\left( {{y}_{1}}\left[ j \right],\ldots ,{{y}_{d}}\left[ j \right] \right)}^{T}}, \\ 
F\left( \mathbf{d} \right)\left[ j \right]&={{\left( F\left( {{d}_{1}} \right)\left[ j \right],\ldots ,F\left( {{d}_{l}} \right)\left[ j \right] \right)}^{T}}, \\ 
F\left( \Delta  \right)\left[ j \right]&={{\left( F\left( {{\Delta }_{1}} \right)\left[ j \right],\ldots ,F\left( {{\Delta }_{l}} \right)\left[ j \right] \right)}^{T}}.  
\end{split}
\end{equation} 
The squared magnitude $\tau ={{\left\| F\left( \mathbf{d} \right)\left[ j \right] \right\|}^{2}}$ is an exponentially distributed variable, with density $f\left( \tau  \right)=\left( {1}/{2\sigma _{\omega }^{2n}}\; \right){{e}^{{-\tau }/{2\sigma _{\omega }^{2}}\;}}$, and from the Parseval theorem \cite{ref17,ref18,ref19} follows, that ${\rm {\mathbb{E}}}\left[\tau \right]\le n2\sigma _{\omega }^{2} $, while the average quadrature modulation variance of the Gaussian subcarriers is
\begin{equation} \label{ZEqnNum999361} 
\sigma _{\omega }^{2} ={\textstyle\frac{1}{n}} \sum _{i=1}^{n}\sigma _{\omega _{i} }^{2}  =\sigma _{\omega _{0} }^{2} . 
\end{equation} 
Eve's attack on sub-channel ${\rm {\mathcal N}}_{i} $ is modeled by the $T_{Eve,i} $ normalized complex transmittance $T_{Eve,i} =\text{Re}T_{Eve,i} +{\rm i}\text{Im}T_{Eve,i} \in {\rm {\mathcal C}}$, where $0\le \text{Re}T_{Eve,i} \le {1 \mathord{\left/{\vphantom{1 \sqrt{2} }}\right.\kern-\nulldelimiterspace} \sqrt{2} } $, $0\le \text{Im}T_{Eve,i} \le {1 \mathord{\left/{\vphantom{1 \sqrt{2} }}\right.\kern-\nulldelimiterspace} \sqrt{2} } $.

\subsection{AMQD-MQA Settings}
Here, we propose the AMQD-MQA for $K\to K$ scenario, that is, for $K$ senders and $K$ receivers. The $1\to K$ and $K\to 1$ cases trivially result from $K\to K$.

\subsubsection{Single Transmitter and Multiple Receivers}
In the first $K\to K$ scheme, a single transmitter generates the input messages of the $K$ independent users. The scheme is based on the AMQD modulation and its sub-channel allocation mechanism \cite{ref4}. The aim of the $K$ independent users is to provide a simultaneous reliable transmission for $K$ independent receivers through the physical Gaussian quantum channel ${\rm {\mathcal N}}$. The multiple access communication is realized by the AMQD modulation, which granulates the inputs of the users into several Gaussian subcarrier CVs. These Gaussian subcarrier CVs are then transmitted through the ${\rm {\mathcal N}}_{i} $ Gaussian sub-channels, following the steps of AMQD. The subset ${\rm {\mathcal A}}$ of transmit users is selected via the procedure of rate selection (see \sref{rate}) at the ${\rm {\mathcal E}}$ encoder. Each ${\rm {\mathcal N}}_{i} $ is allocated by a constant modulation variance $\sigma _{\omega }^{2} $ per the $x$ and $p$ quadrature components, which provably provide an optimal solution in low-SNR regimes because its performance is very close to the exact allocation \cite{ref4,ref20,ref21,ref22}. The Gaussian quadratures that sent via AMQD modulation are dedicated to $K$ independent users. 

\begin{center}
\begin{figure*}[!h]
\begin{center}
\includegraphics[angle = 0,width=1\linewidth]{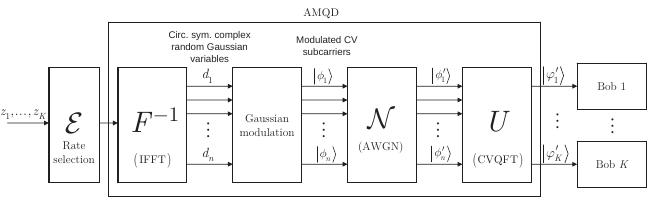}
\caption{The AMQD-MQA multiple access scheme with a single transmitter and multiple receivers. The encoder (${\rm {\mathcal E}}$) modulates and transmits the subcarriers of the $K$ independent users by an AMQD modulation. In the rate-selection phase, Alice selects the users for the transmission and the $\sigma _{\omega _{0,k} }^{2} $ initial modulation variance (per quadrature components) of variable $z_{k} $. The data of the transmit users are then fed into the ${\rm IFFT}$ (inverse fast Fourier transform) operation. The Gaussian subcarrier CVs are transmitted through the good Gaussian sub-channels with a constant modulation variance $\sigma _{\omega }^{2} $. At Bob $k$, each recovered ${\left| \varphi '_{k}  \right\rangle} $ belongs to an independent receiver. The subcarrier CVs are transmitted through $l$ sub-channels, with a total constraint $l\sigma _{\omega }^{2} $ per quadrature components.} 
 \label{fig1}
 \end{center}
\end{figure*}
\end{center}

The first setting of $K\to K$ AMQD-MQA is summarized in \fref{fig1}.

Note that although the main purpose is to improve the performance of the quantum-level transmission in case of AMQD modulation, the aim is different in the AMQD-MQA multiple access scenario. The task is to achieve simultaneous reliable communication among $K$ independent users over a noisy Gaussian quantum channel ${\rm {\mathcal N}}$.

\subsubsection{Multiple Transmitters and Multiple Receivers}

In the second setting of the $K\to K$ model, the system consists of $K$ independent users who transmit the single-carrier Gaussian CVs to an encoder ${\rm {\mathcal E}}$. The encoder applies the rate selection and the inverse CVQFT (continuous-variable quantum Fourier transform, ${\text{CVQFT}}^{\dag } $) operation, which outputs the Gaussian subcarrier CVs. The users select the modulation variances of the $x$ and $p$ quadrature components of the ${\left| \varphi _{k}  \right\rangle} $ single-carrier Gaussian CVs to $\omega _{0,k} $ such that the ${\left| \phi _{i}  \right\rangle} $ Gaussian subcarrier CVs will have variance $\sigma _{\omega }^{2} $ per quadrature components. The transmission of the Gaussian CV subcarriers is realized through $l$ Gaussian sub-channels. The transmitted noisy ${\left| \varphi '_{k}  \right\rangle} $ is received by Bob $k$. 

\begin{center}
\begin{figure*}[!h]
\begin{center}
\includegraphics[angle = 0,width=1\linewidth]{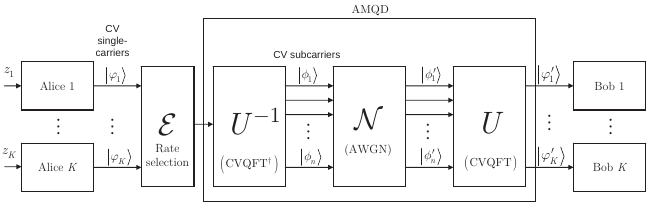}
\caption{The AMQD-MQA multiple access scheme with multiple independent transmitters and multiple receivers. The modulated Gaussian CV single carriers are transformed by a unitary operation (inverse CVQFT) at the ${\rm {\mathcal E}}$ encoder, which outputs the $n$ Gaussian subcarrier CVs for the transmission. The parties send the ${\left| \varphi _{k}  \right\rangle} $ single-carrier Gaussian CVs with variance $\sigma _{\omega _{0,k} }^{2} $ to Alice. In the rate-selection phase, the encoder determines the transmit users. The data states of the transmit users are then fed into the ${\text{CVQFT}}^{\dag } $ operation. The ${\left| \phi _{i}  \right\rangle} $ Gaussian subcarrier CVs have variance $\sigma _{\omega }^{2} $ per quadrature components. The Gaussian CVs are decoded by the CVQFT unitary operation. Each ${\left| \varphi '_{k}  \right\rangle} $ is received by Bob $k$. The subcarrier CVs are transmitted through $l$ sub-channels with a total constraint $l\sigma _{\omega }^{2} $ per quadrature components.} 
 \label{fig2}
 \end{center}
\end{figure*}
\end{center}

The second setting of $K\to K$ AMQD-MQA is summarized in \fref{fig2}.

The rate-selection phase determines the subset ${\rm {\mathcal A}}\subset K$ of transmit users. This procedure is summarized next.

\subsection{Rate Selection}
\label{rate}
The rate $R_{k} $ of user $U_{k} $ can be varied adaptively by the selected number of Gaussian subcarrier CVs that are allocated to user $U_{k} $. The rate selection is defined by an $l\times K$ rate allocation matrix $\mathbf{M}_{\mathcal{A}}^{\left[ j \right]}$ for the $j$-th AMQD block, which allocates the $l$ sub-channels to set ${\rm {\mathcal A}}$ of transmit users. The selected users of ${\rm {\mathcal A}}$ can transmit their quadratures through the $l$ Gaussian sub-channels, in the form of modulated Gaussian subcarrier CVs, $d_{i} \in {\rm {\mathcal C}{\mathcal{N}}}\left(0,\sigma _{d_{i} }^{2} \right)$, $x_{d_{i} } ,p_{d_{i} } \in {\rm {\mathcal{N}}}\left(0,\sigma _{\omega _{i} }^{2} \right)$ (The $i$-th subcarrier $d_{i} $ is sent through the $i$-th ${\rm {\mathcal N}}_{i} $ sub-channel with total variance $2\sigma _{\omega _{i} }^{2} $).

The rate-selection matrix is given in \eqref{matrix} . The matrix $\mathbf{M}_{\mathcal{A}}^{\left[ j \right]}$ identifies the user-sub-channel allocation. If a given $\left(U_{k} ,{\rm {\mathcal N}}_{i} \right)$ element of $\mathbf{M}_{\mathcal{A}}^{\left[ j \right]}$ is 0, then user $U_{k} $ cannot send the quadratures through the Gaussian sub-channel ${\rm {\mathcal N}}_{i} $. If $\left(U_{k} ,{\rm {\mathcal N}}_{i} \right)=1$, then $U_{k} $ can include the transmit data in subcarrier $d_{i} $. The 1s that belong to a given user $U_{k} $ determine the transmission rate of the given user, which is thus the number of Gaussian subcarrier CVs on which $U_{k} $ can send information. If the column of user $U_{k} $ does not contain value 1 in the $l$ rows of $\mathbf{M}_{\mathcal{A}}^{\left[ j \right]}$, then $U_{k} $ does not belong to the subset ${\rm {\mathcal A}}$ of allocated users; thus, $U_{k} $ cannot send information in the $j$-th AMQD block.

In the $\mathbf{M}_{\mathcal{A}}^{\left[ j \right]}$ rate-selection matrix each row defines a subset of ${\rm {\mathcal A}}_{i} \subset K$ from the $K$ users, which identifies the transmit users for the $i$-th subcarrier. The $l$ rows identify an AMQD block. User $U_{k} $ can transmit on sub-channel ${\rm {\mathcal N}}_{i} $ only if $\left(U_{k} ,{\rm {\mathcal N}}_{i} \right)=1$.
\begin{equation} \label{matrix}
\mathbf{M}_{\mathcal{A}}^{\left[ j \right]}=\left( \begin{matrix}
   1 & 0 & \ldots  & 1 & 0  \\
   0 & 1 & \ldots  & 1 & 1  \\
   1 & 0 & \ldots  & 0 & 1  \\
   \vdots  & {} & \ddots  & {} & \vdots   \\
   0 & 1 & \ldots  & 0 & 0  \\
   1 & 1 & \ldots  & 1 & 1  \\
\end{matrix} \right)
\end{equation}

The maximized transmission rate of a single user $U_{k} $ that can be achieved by the rate allocation mechanism, as shown in Equation \eqref{ZEqnNum269002}; for the maximized symmetric rate, see Equation \eqref{ZEqnNum201124}.

\subsection{Entropic Quantities of Gaussian CVs}

Before we provide the steps of the MQA mechanism, it is required to briefly summarize the entropic quantities for the further calculations. 

Let $x$ be a CV variable with probability density function $F_{x} $. The differential entropy \cite{ref17,ref18,ref19}, $H_{{\rm diff}} $, of the $x$ CV variable is as follows:
\begin{equation} \label{ZEqnNum203317} 
H_{{\rm diff}} \left(x\right)=\int\limits_{-\infty }^{\infty }F_{x}  \left(u\right)\log _{2} \left({\textstyle\frac{1}{F_{x} \left(u\right)}} \right){\rm d}u. 
\end{equation} 
The conditional differential entropy, at given output $y$ is
\begin{equation} \label{31)} 
H_{{\rm diff}} \left(\left. x\right|y\right)=\int\limits_{-\infty }^{\infty }F_{x,y}  \left(u,v\right)\log _{2} \left({\textstyle\frac{1}{F_{\left. x\right|y} \left(\left. u\right|v\right)}} \right){\rm d}u{\rm d}v. 
\end{equation} 
From these quantities, the continuous mutual information function is
\begin{equation} \label{ZEqnNum422557} 
I\left(x:y\right)=H_{{\rm diff}} \left(x\right)-H_{{\rm diff}} \left(\left. x\right|y\right),                                 
\end{equation} 
whereas the capacity of a continuous Gaussian quantum channel ${\rm {\mathcal N}}$ (also referred to as AWGN - additive white Gaussian noise channel) is the maximization of Equation \eqref{ZEqnNum422557} with respect to the density function $F_{x} $ of $x$:
\begin{equation} \label{33)} 
C\left({\rm {\mathcal N}}\right)=\mathop{\max }\limits_{F_{x} } I\left(x:y\right).                                           
\end{equation} 
Assuming a given power constraint ${\textstyle\frac{1}{N}} \sum _{N}\sigma _{\omega _{i} }^{2}  \le \sigma _{\omega }^{2} $, $\sigma _{x_{i} }^{2} ={\rm {\mathbb{E}}}\left[c\left(x\right)\right]\le \sigma _{\omega }^{2} $, where $c\left(\cdot \right)$ is the cost function, the constrained capacity is
\begin{equation} \label{34)} 
C\left({\rm {\mathcal N}}\right)=\mathop{\max }\limits_{F_{x} :{\rm {\mathbb{E}}}\left[c\left(x\right)\right]\le \sigma _{\omega }^{2} } I\left(x:y\right).                                       
\end{equation} 
Assume that the Gaussian CV state is modeled by $g$, a zero-mean Gaussian random variable $g\in {\rm {\mathcal{N}}}\left(0,\sigma _{\omega }^{2} \right)$. The cost function \cite{ref17} of the Gaussian continuous channel is $c\left(g\right)=g^{2} $, whereas for a zero-mean, circular symmetric complex Gaussian random variable $z\in {\rm {\mathcal C}{\mathcal{N}}}\left(0,\sigma _{z}^{2} \right)$, $\sigma _{z}^{2} ={\rm {\mathbb{E}}}\left[\left|z\right|^{2} \right]$, $c\left(z\right)=\left|z\right|^{2} $.

The differential entropy of $g\in {\rm {\mathcal{N}}}\left(0,\sigma _{\omega }^{2} \right)$ is
\begin{equation} \label{35)} 
H_{{\rm diff}} \left(g\right)={\textstyle\frac{1}{2}} \log _{2} \left(2\pi e\sigma _{\omega }^{2} \right),                                       
\end{equation} 
whereas the conditional differential entropy is 
\begin{equation} \label{36)} 
H_{{\rm diff}} \left(\left. g\right|y\right)={\textstyle\frac{1}{2}} \log _{2} \left(2\pi e\sigma _{{\rm {\mathcal N}}}^{2} \right),                                      
\end{equation} 
where $y=g+\Delta $, $\Delta \in {\rm {\mathcal{N}}}\left(0,\sigma _{{\rm {\mathcal N}}}^{2} \right)$ is the zero-mean Gaussian noise of the quantum channel.

The differential entropy of $z\in {\rm {\mathcal C}{\mathcal{N}}}\left(0,\sigma _{z}^{2} \right)$ is 
\begin{equation} \label{37)} 
\begin{split}
   {{H}_{\text{diff}}}\left( z \right)&={{H}_{\text{diff}}}\left( \operatorname{\text{Re}}\left( z \right) \right)+{{H}_{\text{diff}}}\left( \operatorname{\text{Im}}\left( z \right) \right) \\ 
 & =\tfrac{1}{2}{{\log }_{2}}\left( 2\pi e\sigma _{z}^{2} \right)=\tfrac{1}{2}{{\log }_{2}}\left( 2\pi e2\sigma _{{{\omega }_{0}}}^{2} \right),  
\end{split}
\end{equation} 
whereas the conditional differential entropy of $z$ is
\begin{equation} \label{38)} 
\begin{split}
  {{H}_{\text{diff}}}\left( \left. z \right|y \right)&={{H}_{\text{diff}}}\left( \operatorname{\text{Re}}\left( \left( \left. z \right|y \right) \right) \right)+{{H}_{\text{diff}}}\left( \operatorname{\text{Im}}\left( \left( \left. z \right|y \right) \right) \right) \\ 
 & =\tfrac{1}{2}{{\log }_{2}}\left( 2\pi e\sigma _{\Delta }^{2} \right)=\tfrac{1}{2}{{\log }_{2}}\left( 2\pi e2\sigma _{\mathcal{N}}^{2} \right),  
\end{split}
\end{equation} 
because the differential entropies are invariant to translations of the density \cite{ref17}. 

The continuous mutual information for the Gaussian random distribution is
\begin{equation} \label{39)} 
I\left(g:y\right)=H_{{\rm diff}} \left(y\right)-H_{{\rm diff}} \left(\left. y\right|g\right)=H_{{\rm diff}} \left(y\right)-{\textstyle\frac{1}{2}} \log _{2} \left(2\pi e\sigma _{{\rm {\mathcal N}}}^{2} \right),               
\end{equation} 
whereas the constrained capacity of the Gaussian quantum channel for an input $g$ is
\begin{equation} \label{ZEqnNum877787} 
C\left({\rm {\mathcal N}}\right)=\mathop{\max }\limits_{F_{x} :{\rm {\mathbb{E}}}\left[g^{2} \right]\le \sigma _{\omega }^{2} } I\left(x:y\right).                                          
\end{equation} 
Equation \eqref{ZEqnNum877787} is justified by the fact that the CV input and output can be discretized, and the continuous channel can be approximated as a discrete channel. 

Assuming a ${\rm {\mathcal C}{\mathcal{N}}}\left(0,\sigma _{z}^{2} \right)$ distribution, the constrained capacity for an input $z$ drawn from this distribution is
\begin{equation} \label{41)} 
C\left({\rm {\mathcal N}}\right)=\mathop{\max }\limits_{F_{x} :{\rm {\mathbb{E}}}\left[\left|z\right|^{2} \right]\le 2\sigma _{\omega _{0} }^{2} } I\left(x:y\right).                                      
\end{equation} 
A zero-mean Gaussian random variable maximizes the entropy; hence, they are also maximally capacity constraint satisfiers. The constraint of ${\rm {\mathbb{E}}}\left[x^{2} \right]\le \sigma _{\omega }^{2} $ on a random variable $x$ is satisfied by the $g\in {\rm {\mathcal{N}}}\left(0,\sigma _{\omega }^{2} \right)$ and the constraint of ${\rm {\mathbb{E}}}\left[\left|x\right|^{2} \right]\le 2\sigma _{\omega _{0} }^{2} $ is satisfied by a $x=z\in {\rm {\mathcal C}{\mathcal{N}}}\left(0,\sigma _{z}^{2} \right),\sigma _{z}^{2} ={\rm {\mathbb{E}}}\left[\left|z\right|^{2} \right]$ distributed random variable with i.i.d. Gaussian random real and imaginary parts $\text{Re},\text{Im}\in {\rm {\mathcal{N}}}\left(0,\sigma _{\omega _{0} }^{2} \right)$. This leads to the following real-domain capacity of ${\rm {\mathcal N}}$:
\begin{equation} \label{42)} 
C\left({\rm {\mathcal N}}\right)={\textstyle\frac{1}{2}} \log _{2} \left(2\pi e\left(\sigma _{\omega }^{2} +\sigma _{{\rm {\mathcal N}}}^{2} \right)\right)-{\textstyle\frac{1}{2}} \log _{2} \left(2\pi e\sigma _{{\rm {\mathcal N}}}^{2} \right)={\textstyle\frac{1}{2}} \log _{2} \left(1+{\textstyle\frac{\sigma _{\omega }^{2} }{\sigma _{{\rm {\mathcal N}}}^{2} }} \right).          
\end{equation} 
For a $z\in {\rm {\mathcal C}{\mathcal{N}}}\left(0,\sigma _{z}^{2} \right),\sigma _{z}^{2} ={\rm {\mathbb{E}}}\left[\left|z\right|^{2} \right]=2\sigma _{\omega _{0} }^{2} $ input, with i.i.d. Gaussian noise components on the $x$ position and $p$ momentum quadratures, and complex noise $\Delta \in {\rm {\mathcal C}{\mathcal{N}}}\left(0,\sigma _{\Delta }^{2} \right)=\Delta _{x} +{\rm i}\Delta _{p} $, with $\Delta _{x} ,\Delta _{p} \in {\rm {\mathcal{N}}}\left(0,\sigma _{{\rm {\mathcal N}}}^{2} \right)$, the complex-domain capacity of ${\rm {\mathcal N}}$ is
\begin{equation} \label{43)} 
\begin{split}
   C\left( \mathcal{N} \right)&=\tfrac{1}{2}{{\log }_{2}}\left( 2\pi e\left( \sigma _{z}^{2}+\sigma _{\Delta }^{2} \right) \right)-\tfrac{1}{2}{{\log }_{2}}\left( 2\pi e\sigma _{\Delta }^{2} \right) \\ 
 & =\tfrac{1}{2}{{\log }_{2}}\left( 2\pi e\left( 2\sigma _{{{\omega }_{0}}}^{2}+2\sigma _{\mathcal{N}}^{2} \right) \right)-\tfrac{1}{2}{{\log }_{2}}\left( 2\pi e2\sigma _{\mathcal{N}}^{2} \right) \\ 
 & ={{\log }_{2}}\left( 1+\tfrac{2\sigma _{{{\omega }_{0}}}^{2}}{2\sigma _{\mathcal{N}}^{2}} \right)={{\log }_{2}}\left( 1+\tfrac{\sigma _{{{\omega }_{0}}}^{2}}{\sigma _{\mathcal{N}}^{2}} \right),  
\end{split}
\end{equation} 
whereas the real-domain capacity (i.e., with respect to a given quadrature component) is
\begin{equation} \label{44)} 
\begin{split}
   C\left( \mathcal{N} \right)&=\tfrac{1}{2}{{\log }_{2}}\left( 2\pi e\left( \tfrac{1}{2}\sigma _{z}^{2}+\tfrac{1}{2}\sigma _{\Delta }^{2} \right) \right)-\tfrac{1}{2}{{\log }_{2}}\left( 2\pi e\tfrac{1}{2}\sigma _{\Delta }^{2} \right) \\ 
 & =\tfrac{1}{2}{{\log }_{2}}\left( 2\pi e\left( \sigma _{\omega }^{2}+\sigma _{\mathcal{N}}^{2} \right) \right)-\tfrac{1}{2}{{\log }_{2}}\left( 2\pi e\sigma _{\mathcal{N}}^{2} \right) \\ 
 & =\tfrac{1}{2}{{\log }_{2}}\left( 1+\tfrac{2\sigma _{{{\omega }_{0}}}^{2}}{2\sigma _{\mathcal{N}}^{2}} \right)=\tfrac{1}{2}{{\log }_{2}}\left( 1+\tfrac{\sigma _{{{\omega }_{0}}}^{2}}{\sigma _{\mathcal{N}}^{2}} \right).  
\end{split}
\end{equation} 
In using the AMQD-MQA, the $d\in {\rm {\mathcal C}{\mathcal{N}}}\left(0,\sigma _{d}^{2} \right),\sigma _{d}^{2} ={\rm {\mathbb{E}}}\left[\left|d\right|^{2} \right]=2\sigma _{\omega }^{2} $ Gaussian subcarrier CV states, more precisely the $\sigma _{\omega }^{2} $ modulation variance of these subcarriers, $x_{d} ,p_{d} \in {\rm {\mathcal{N}}}\left(0,\sigma _{\omega }^{2} \right)$, will formulate the achievable capacity.

Note that throughout the manuscript, the complex-domain capacity formula will be used, and the modulation variance allocation will be discussed with respect to the independent position ($x$) and momentum ($p$) quadrature components of the Gaussian CV states.

\section{AMQD-MQA Multiple Access}
\label{sec3}
\begin{theorem}
(Multiple access capacity-achieving communication in AMQD-MQA). In an AMQD-MQA multiple access scheme with $K$ independent users, the sum rate over the $l$ sub-channels at a constant $\sigma _{\omega }^{2} $ for each ${\rm {\mathcal N}}_{i} $ is $R_{sum}^{MQA} \le \mathop{\max }\limits_{\forall i} \sum _{l}\log _{2} \left(1+\sigma _{\omega _{i} }^{2} {\left|F\left(T_{i} \left({\rm {\mathcal N}}_{i} \right)\right)\right|^{2}  \mathord{\left/{\vphantom{\left|F\left(T_{i} \left({\rm {\mathcal N}}_{i} \right)\right)\right|^{2}  \sigma _{{\rm {\mathcal N}}}^{2} }}\right.\kern-\nulldelimiterspace} \sigma _{{\rm {\mathcal N}}}^{2} } \right) ,$ whereas the symmetric rate is $R_{sym}^{MQA} \le {\textstyle\frac{1}{K}} \mathop{\max }\limits_{\forall i} \sum _{l}\log _{2} \left(1+{\sigma _{\omega _{i} }^{2} \left|F\left(T_{i} \left({\rm {\mathcal N}}_{i} \right)\right)\right|^{2}  \mathord{\left/{\vphantom{\sigma _{\omega _{i} }^{2} \left|F\left(T_{i} \left({\rm {\mathcal N}}_{i} \right)\right)\right|^{2}  \sigma _{{\rm {\mathcal N}}}^{2} }}\right.\kern-\nulldelimiterspace} \sigma _{{\rm {\mathcal N}}}^{2} } \right) ,$ where $\left|F\left(T_{i} \right)\right|^{2} $ is the squared magnitude of the CVQFT-transformed $T_{i} $ transmittance coefficient of the $i$-th Gaussian sub-channel ${\rm {\mathcal N}}_{i} $.
\end{theorem}
\begin{proof}
The proof demonstrates the results for $K=2$, that is, for two independent users, $U_{1} $ and $U_{2} $. The proof is organized as follows. In the first part, the ${\rm{C}}$ capacity region of a multiple access quantum channel is derived. In the second part, we add the CVQFT operation into the picture and revise the results of the first part, which then leads to the final rates of the users. Let the inputs of user $U_{k} $, $k=1,\ldots ,K$ to be modeled by the zero-mean, circular symmetric variables ${{z}_{k}}={{x}_{k}}+i{{p}_{k}}\in \mathcal{C}\mathbb{N}\left( 0,\sigma _{{{z}_{k}}}^{2} \right)$, $\sigma _{{{z}_{k}}}^{2}=\mathbb{E}\left[ {{\left| {{z}_{k}} \right|}^{2}} \right]$, $x_{k} \in {\rm {\mathcal{N}}}\left(0,\sigma _{\omega _{k,0} }^{2} \right)$, $p_{k} \in {\rm {\mathcal{N}}}\left(0,\sigma _{\omega _{k,0} }^{2} \right)$. Let the $i$-th Gaussian subcarrier CV be $d_{i} =x_{d_{i} } +{\rm i}p_{d_{i} } \in {\rm {\mathcal C}{\mathcal{N}}}\left(0,\sigma _{d_{i} }^{2} \right)$, $\sigma _{d_{i} }^{2} ={\rm {\mathbb{E}}}\left[\left|d_{i} \right|^{2} \right]$ with i.i.d. Gaussian random quadratures $x_{d_{i} } \in {\rm {\mathcal{N}}}\left(0,\sigma _{\omega _{i} }^{2} \right)$, $p_{d_{i} } \in {\rm {\mathcal{N}}}\left(0,\sigma _{\omega _{i} }^{2} \right)$. 

Let $l$ be the number of good ${\rm {\mathcal N}}_{i} $ Gaussian sub-channels. (For an exact clarification, it precisely refers to the following. The noise of the sub-channels is below the critical security parameter $\nu _{{\text{Eve}}} $, which identifies the optimal Gaussian collective attack, see Equation \eqref{ZEqnNum791007} and \cite{ref4}). Let the transmittance of the $i$-th sub-channel be $T_{i} \left({\rm {\mathcal N}}_{i} \right)\in {\rm {\mathcal C}}$. 

The outputs of $U_{1} $ and $U_{2} $ are expressed as follows:
\begin{equation} \label{45)} 
y_{k} =T\left({\rm {\mathcal N}}\right)z_{k} +\Delta ,{\rm \; }k=1,2.{\rm \; } 
\end{equation} 
The $d$-dimensional output of the $k$-th user is
\begin{equation} \label{ZEqnNum831044} 
{{\mathbf{y}}_{k}}=\mathbf{T}\left( \mathcal{N} \right){{\mathbf{z}}_{k}}+\Delta ,
\end{equation} 
where $\mathbf{T}\left( \mathcal{N} \right)={{\left[ {{T}_{k,1}}\left( \mathcal{N} \right),\ldots ,{{T}_{k,d}}\left( \mathcal{N} \right) \right]}^{T}}$, ${{\mathbf{z}}_{k}}\left( \mathcal{N} \right)={{\left[ {{z}_{k,1}},\ldots ,{{z}_{k,d}} \right]}^{T}}\in \mathcal{C}\mathcal{N}\left( 0,{{\mathbf{K}}_{{{\mathbf{z}}_{k}}}} \right)$, and ${{\mathbf{K}}_{{{\mathbf{z}}_{k}}}}$ is the covariance matrix of the zero-mean, circular symmetric Gaussian random vector ${{\mathbf{z}}_{k}}\in \mathcal{C}\mathcal{N}\left( 0,{{\mathbf{K}}_{{{\mathbf{z}}_{k}}}} \right)$ of $U_{k} $. 

The sum capacity \cite{ref17,ref18,ref19} is the total throughput over the $l$ sub-channels of ${\rm {\mathcal N}}$ at a constant modulation variance $\sigma _{\omega }^{2} $ is as follows: 
\begin{equation} \label{ZEqnNum842397} 
\begin{split}
   {{C}_{\text{sum}}}\left( \mathcal{N} \right)&=\underset{\left( {{R}_{1}},{{R}_{2}} \right)\in \rm{C}}{\mathop{\max }}\,{{R}_{1}}+{{R}_{2}} \\ 
 & =\underset{\forall i}{\mathop{\max }}\,\sum\nolimits_{l}{{{\log }_{2}}\left( 1+\tfrac{\sigma _{{{\omega }_{i}}}^{2}{{\left| {{T}_{i}}\left( {{\mathcal{N}}_{i}} \right) \right|}^{2}}}{\sigma _{\mathcal{N}}^{2}} \right)}.  
\end{split}
\end{equation} 
The symmetric capacity \cite{ref17,ref18,ref19} is the maximum common rate at which both $U_{1} $ and $U_{2} $ can reliably transmit information over the $l$ sub-channels of ${\rm {\mathcal N}}$, as follows:
\begin{equation} \label{48)} 
C_{{\rm sym}} \left({\rm {\mathcal N}}\right)=\mathop{\max }\limits_{\left(R_{{\rm sym}} ,R_{{\rm sym}} \right)\in {\rm{C}}} R_{{\rm sym}} ={\textstyle\frac{1}{2}} \mathop{\max }\limits_{\forall i} \sum _{l}\log _{2} \left(1+{\textstyle\frac{\sigma _{\omega _{i} }^{2} \left|T_{i} \left({\rm {\mathcal N}}_{i} \right)\right|^{2} }{\sigma _{{\rm {\mathcal N}}}^{2} }} \right) , 
\end{equation} 
where $R_{{\rm sym}} $ is the rate at which both $U_{1} $ and $U_{2} $ can simultaneously communicate in a reliable form.

For $K$ users $U_{1} ,\ldots U_{K} $, the sum capacity and the symmetric capacity of ${\rm {\mathcal N}}$ are expressed as 
\begin{equation} \label{49)} 
C_{{\rm sum}} \left({\rm {\mathcal N}}\right)=\mathop{\max }\limits_{\left(R_{1} ,\ldots ,R_{K} \right)\in {\rm{C}}} \sum _{K}R_{i}  =\mathop{\max }\limits_{\forall i} \sum _{l}\log _{2} \left(1+{\textstyle\frac{\sigma _{\omega _{i} }^{2} \left|T_{i} \left({\rm {\mathcal N}}_{i} \right)\right|^{2} }{\sigma _{{\rm {\mathcal N}}}^{2} }} \right) , 
\end{equation} 
and
\begin{equation} \label{50)} 
C_{{\rm sym}} \left({\rm {\mathcal N}}\right)=\mathop{\max }\limits_{\left(R_{{\rm sym}} ,\ldots ,R_{{\rm sym}} \right)\in {\rm{C}}} R_{{\rm sym}} ={\textstyle\frac{1}{K}} \mathop{\max }\limits_{\forall i} \sum _{l}\log _{2} \left(1+{\textstyle\frac{\sigma _{\omega _{i} }^{2} \left|T_{i} \left({\rm {\mathcal N}}_{i} \right)\right|^{2} }{\sigma _{{\rm {\mathcal N}}}^{2} }} \right) .            
\end{equation} 
The ${\rm{C}}$ capacity region \cite{ref17,ref18,ref19} is the region of the rates of $\left(R_{1} ,R_{2} \right)$ of $U_{1} $ and $U_{2} $, at which both users can have a simultaneous reliable communication over the quantum channel ${\rm {\mathcal N}}$. The region ${\rm{C}}$ upper bounds the independent single transmission rates as of $U_{1} $ and $U_{2} $, which can be maximized if all the $l$ sub-channels with a total constraint $l\sigma _{\omega }^{2} $ (i.e., all degrees of freedom) are dedicated to user $k$, as follows:
\begin{equation} \label{51)} 
R_{k} \le \mathop{\max }\limits_{\forall i} \sum _{l}\log _{2} \left(1+{\textstyle\frac{\sigma _{\omega _{i} }^{2} \left|T_{i} \left({\rm {\mathcal N}}_{i} \right)\right|^{2} }{\sigma _{{\rm {\mathcal N}}}^{2} }} \right) ,{\rm \; }k=1,\ldots ,K.                      
\end{equation} 
If the equality holds, then only user $U_{k} $ is allowed to transmit over the $l$ sub-channels. These rates define the corner points $C_{1} =\mathop{\max }\limits_{\left(R_{1} \right)\in {\rm {\mathcal C}}} R_{1} $ and $C_{2} =\mathop{\max }\limits_{\left(R_{2} \right)\in {\rm {\mathcal C}}} R_{2} $ of $U_{1} $ and $U_{2} $. At the corner points, the rate of the given user is maximal whereas rate of the other user is zero. On the line between the corner points, both users are allowed to simultaneously communicate at rates $R_{1} $ and $R_{2} $. 

The sum rate $R_{1} +R_{2} $ is defined by a line between the corner points $C_{1} $ and $C_{2} $. The sum rate is bounded by Equation \eqref{ZEqnNum842397}; hence, 
\begin{equation} \label{52)} 
R_{1} +R_{2} \le \mathop{\max }\limits_{\forall i} \sum _{l}\log _{2} \left(1+{\textstyle\frac{\sigma _{\omega _{i} }^{2} \left|T_{i} \left({\rm {\mathcal N}}_{i} \right)\right|^{2} }{\sigma _{{\rm {\mathcal N}}}^{2} }} \right) . 
\end{equation} 
The ${\rm{C}}$ capacity region can be expressed in terms of the mutual information $I\left(z_{k} :y_{k} ,T\left({\rm {\mathcal N}}\right)\right)$ of the Gaussian random inputs $z_{k} =x_{1} +{\rm i}p_{1} \in {\rm {\mathcal C}{\mathcal{N}}}\left(0,{\rm {\mathbb{E}}}\left[\left|z_{k} \right|^{2} \right]\right)$, and channel output
\begin{equation} \label{53)} 
\begin{split}
   y&={{y}_{1}}+{{y}_{2}} \\ 
 & =T\left( \mathcal{N} \right){{z}_{1}}+{{\Delta }_{1}}+T\left( \mathcal{N} \right){{z}_{2}}+{{\Delta }_{2}}.  
\end{split}	   
\end{equation} 
In the ${\rm{C}}\left(z_{1} ,z_{2} \right)$ capacity region for this Gaussian input distribution, the rate pairs are
\begin{equation} \label{ZEqnNum803213} 
\begin{split}
  & {{R}_{1}}\le I\left( {{z}_{1}}:\left. y,T\left( \mathcal{N} \right) \right|{{z}_{2}} \right), \\ 
 & {{R}_{2}}\le I\left( {{z}_{2}}:\left. y,T\left( \mathcal{N} \right) \right|{{z}_{1}} \right), \\ 
\end{split}
\end{equation} 
and
\begin{equation} \label{ZEqnNum579366} 
R_{1} +R_{2} \le I\left(z_{1} ,z_{2} :y,T\left({\rm {\mathcal N}}\right)\right).                           
\end{equation} 
From the chain rule of mutual information,
\begin{equation} \label{56)} 
I\left(z_{1} ,z_{2} :y\right)=I\left(z_{1} :y\right)+I\left(z_{2} :\left. y\right|z_{1} \right).                        
\end{equation} 
One immediately can conclude that
\begin{equation} \label{57)} 
\begin{split}
   I\left( {{z}_{k}}:{{y}_{k}},T\left( \mathcal{N} \right) \right)&=I\left( {{z}_{k}}:T\left( \mathcal{N} \right) \right)+I\left( {{z}_{k}}:\left. y \right|T\left( \mathcal{N} \right) \right) \\ 
 & =I\left( {{z}_{k}}:\left. y \right|T\left( \mathcal{N} \right) \right).  
\end{split}
\end{equation} 
Because $I\left(z_{k} :y_{k} ,T\left({\rm {\mathcal N}}\right)\right)$ of the $l$ sub-channels is independent of the input $z_{k} $, thus $I\left(z_{k} :T\left({\rm {\mathcal N}}\right)\right)=0$ \cite{ref17,ref18,ref19}.

Therefore, conditioned on $T\left({\rm {\mathcal N}}\right)$, the channel model is perfectly analogous to an AWGN channel, with the following SNR at a given sub-channel ${\rm {\mathcal N}}_{i} $:
\begin{equation} \label{58)} 
{\rm SNR}={\textstyle\frac{\sigma _{\omega _{i} }^{2} }{\sigma _{{\rm {\mathcal N}}_{i} }^{2} }} .                                                
\end{equation} 
It means that the ideal input distribution is the zero-mean, circular symmetric complex Gaussian random inputs $z_{k} =x_{k} +{\rm i}p_{k} \in {\rm {\mathcal C}{\mathcal{N}}}\left(0,{\rm {\mathbb{E}}}\left[\left|z_{k} \right|^{2} \right]\right)$, independent from the actual SNR values. In other words, the mutual information can be maximized by the zero-mean ${\rm {\mathcal C}{\mathcal{N}}}$ distribution, which leads to the following:
\begin{equation} \label{59)} 
I\left(z_{k} :\left. y\right|T\left({\rm {\mathcal N}}\right)\right)=\sum _{l}\log _{2} \left(1+{\textstyle\frac{\sigma _{\omega _{i} }^{2} \left|T_{i} \left({\rm {\mathcal N}}_{i} \right)\right|^{2} }{\sigma _{{\rm {\mathcal N}}}^{2} }} \right) .                  
\end{equation} 
Exploiting the chain rule of mutual information, the capacity regions shown in Equations \eqref{ZEqnNum803213} and \eqref{ZEqnNum579366} can be rewritten as
\begin{equation} \label{60)} 
\begin{split}
  & {{R}_{1}}\le I\left( {{z}_{1}}:\left. y \right|T\left( \mathcal{N} \right),{{z}_{2}} \right), \\ 
 & {{R}_{2}}\le I\left( {{z}_{2}}:\left. y \right|T\left( \mathcal{N} \right),{{z}_{1}} \right), \\ 
\end{split}
\end{equation} 
and
\begin{equation} \label{61)} 
R_{1} +R_{2} \le I\left(z_{1} ,z_{2} :\left. y\right|T\left({\rm {\mathcal N}}\right)\right).                           
\end{equation} 
The corner points $C_{1} $ and $C_{2} $ are defined as follows:
\begin{equation} \label{62)} 
\begin{split}
   {{C}_{1}}&\equiv I\left( {{z}_{1}}:\left. y,T\left( \mathcal{N} \right) \right|{{z}_{2}} \right) \\ 
 & =\underset{\left( {{R}_{1}} \right)\in \rm{C}}{\mathop{\max }}\,{{R}_{1}} \\ 
 & =\underset{\forall i}{\mathop{\max }}\,\sum\nolimits_{l}{{{\log }_{2}}\left( 1+\tfrac{\sigma _{{{\omega }_{i}}}^{2}{{\left| {{T}_{i}}\left( {{\mathcal{N}}_{i}} \right) \right|}^{2}}}{\sigma _{\mathcal{N}}^{2}} \right)},  
\end{split}
\end{equation} 
and
\begin{equation} \label{63)} 
\begin{split}
   {{C}_{2}}&\equiv I\left( {{z}_{2}}:\left. y,T\left( \mathcal{N} \right) \right|{{z}_{1}} \right) \\ 
 & =\underset{\left( {{R}_{2}} \right)\in \rm{C}}{\mathop{\max }}\,{{R}_{2}} \\ 
 & =\underset{\forall i}{\mathop{\max }}\,\sum\nolimits_{l}{{{\log }_{2}}\left( 1+\tfrac{\sigma _{{{\omega }_{i}}}^{2}{{\left| {{T}_{i}}\left( {{\mathcal{N}}_{i}} \right) \right|}^{2}}}{\sigma _{\mathcal{N}}^{2}} \right)}.  
\end{split}
\end{equation} 
Hence, in the corner points $C_{1} $, $C_{2} $, only $U_{1} $, $U_{2} $ is allowed to transmit over the $l$ sub-channels, whereas the rate of the other user is zero. Taking the ${\rm {\mathcal H}}$ convex hull of all possible independent input distributions leads to the capacity region ${\rm{C}}$ as
\begin{equation} \label{ZEqnNum992824} 
{\rm{C}={\mathcal H}}\left(\bigcup _{z_{1} ,z_{2} }{\rm{C}}\left(z_{1} ,z_{2} \right) \right).                                   
\end{equation} 
The inputs $z_{1} $ and $z_{2} $ are zero-mean Gaussian random variables, for which follows that all information quantities that characterize capacity region ${\rm{C}}$ are simultaneously maximized because the capacity region ${\rm{C}}\left({\rm {\mathcal C}{\mathcal{N}}},{\rm {\mathcal C}{\mathcal{N}}}\right)$ with zero-mean, circular symmetric complex Gaussian random distribution variables formulates the superset of all other capacity regions (for the proof of optimality, see Lemma 1) with arbitrary $p_{x'} $ distributions, that is, 
\begin{equation} \label{65)} 
{\rm{C}}\left({\rm {\mathcal C}{\mathcal{N}}},{\rm {\mathcal C}{\mathcal{N}}}\right)={\rm {\mathcal S}}\left(\bigcup _{\forall p_{x'} }{\rm{C}}\left(p_{x'} ,p_{x'} \right) \right)={\rm {\mathcal S}}\bigcup _{\forall p_{x'} }{\rm {\mathcal H}}\left(\bigcup _{x'_{1} ,x'_{2} }{\rm{C}}\left(x'_{1} ,x'_{2} \right) \right) .                   
\end{equation} 
As follows, for the $z_{k} \in {\rm {\mathcal C}{\mathcal{N}}}\left(0,{\rm {\mathbb{E}}}\left[\left|z_{k} \right|^{2} \right]\right)$ inputs, one obtains the capacity region ${\rm{C}}$:
\begin{equation} \label{66)} 
\begin{split}
  & {{I}_{\text{MQA}}}\left( {{z}_{1}}:\left. y \right|{{z}_{2}} \right)=\underset{\forall {{{{x}'_{1}}}}}{\mathop{\max }}\,I\left( {{{{x}'_{1}}}}:\left. y \right|{{{{x}'_{2}}}} \right), \\ 
 & {{I}_{\text{MQA}}}\left( {{z}_{2}}:\left. y \right|{{z}_{1}} \right)=\underset{\forall {{{{x}'_{2}}}}}{\mathop{\max }}\,I\left( {{{{x}'_{2}}}}:\left. y \right|{{{{x}'_{1}}}} \right), \\ 
\end{split}
\end{equation} 
and
\begin{equation} \label{67)} 
I_{{\rm MQA}} \left(z_{1} ,z_{2} :y\right)=\mathop{\max }\limits_{\forall x'_{1} ,x'_{2} } I\left(x'_{1} ,x'_{2} :y\right).                                    
\end{equation} 
Let us now add the CVQFT operation into the picture to derive the final rates.

Assuming a subset ${\rm {\mathcal A}}\subseteq K$ of allowed users in set $K$ with $\left|{\rm {\mathcal A}}\right|$ users (these $\left|{\rm {\mathcal A}}\right|$ users have the value of 1 in the $i$-th row of the rate-selection matrix $\mathbf{M}_{\mathcal{A}}^{\left[ j \right]}$, see \eqref{matrix}),
\begin{equation} \label{68)} 
d_{i} =e^{{\textstyle\frac{\sigma _{\omega _{0} }^{2} d_{i}^{2} }{2}} } .                                               
\end{equation} 
The Gaussian CV subcarrier variable can be rewritten as
\begin{equation} \label{ZEqnNum166801} 
d_{i} ={\textstyle\frac{1}{\sqrt{n} }} \sum _{k=0}^{\left|{\rm {\mathcal A}}\right|-1}z_{k}  e^{\frac{-{\rm i}2\pi ik}{n} } ,i=0,\ldots ,n-1,                            
\end{equation} 
where $n$ is the number of outputs of the Fourier transform (from these $n$ outputs, only $l$ will be used in the subcarrier transmission, by the initial assumption on the conditions of the sub-channels.), $z_{k} $ is the input of user ${\rm {\mathcal A}}\in k$. If for a given user $U_{k} ,{\rm \; }k\in K$, the ${\rm {\mathcal A}}\not\subset k$ relation holds, then user $U_{k} $ cannot transmit data in the subcarrier $d_{i} $, and the input $z_{k} $ of $U_{k} $ is not included in Equation \eqref{ZEqnNum166801}.

In the AMQD-MQA, an output block (for the derivation of an AMQD block, see \cite{ref4}) is characterized as
\begin{equation} \label{ZEqnNum907723} 
{{\mathbf{y}}_{k}}\left[ j \right]=\sum\nolimits_{k=1}^{\left| \mathcal{A} \right|}{F\left( {{T}_{k,i}}\left( {{\mathcal{N}}_{i}} \right) \right)}\left[ j \right]F\left( {{d}_{k,i}} \right)\left[ j \right]+F\left( {{\Delta }_{i}} \right)\left[ j \right],\text{ }i=0\ldots l-1,j=1\ldots \left| \mathcal{A} \right|,
\end{equation} 
where $\left[j\right]$ refers to the $j$-th AMQD block, $k$ identifies user $U_{k} $, $F\left(T_{k,i} \left({\rm {\mathcal N}}_{i} \right)\right)\left[j\right]$ is the CVQFT of the transmittance coefficient of the $i$-th Gaussian sub-channel ${\rm {\mathcal N}}_{i} $, $F\left(d_{k,i} \right)\left[j\right]$ is the CVQFT-transformed subcarrier $d_{k,i} $, and $F\left(\Delta _{i} \right)\left[j\right]$ is the Fourier-transformed Gaussian noise $\Delta _{i} \in {\rm {\mathcal C}{\mathcal{N}}}\left(0,\sigma _{\Delta _{i} }^{2} \right)$, $\sigma _{\Delta _{i} }^{2} ={\rm {\mathbb{E}}}\left[\left|\Delta _{i} \right|^{2} \right]$ of the $i$-th sub-channel. As Equation \eqref{ZEqnNum907723} reveals, in the $i$-th subcarrier $d_{i} $, only the allocated users' data are conveyed. 

The capacity that can be achieved by an AMQD-MQA block is derived as follows. Assuming channel output (Equation \eqref{ZEqnNum907723}) with modulation variance $\sigma _{\omega }^{2} $ per quadrature components of $d_{i} $, from the Parseval theorem, it follows that for $F\left( {{\mathbf{d}}_{k}} \right)\left[ j \right]={{\left[ F\left( {{d}_{k,1}} \right)\left[ j \right],\ldots F\left( {{d}_{k,l}} \right)\left[ j \right] \right]}^{T}}$,
\begin{equation} \label{71)} 
\sigma _{F\left( {{\mathbf{d}}_{k}}\left[ j \right] \right)}^{2}=\mathbb{E}\left[ {{\left\| F\left( {{\mathbf{d}}_{k}} \right)\left[ j \right] \right\|}^{2}} \right]\le 2l\sigma _{\omega }^{2},
\end{equation} 
Introducing ${{\mathbf{y}}_{k}}\left[ j \right]$ leads to capacity (averaged over an AMQD block),
\begin{equation} \label{72)} 
{{C}_{\text{sum}}}\left( \mathcal{N} \right)=\underset{\mathbb{E}\left[ {{\left| F\left( {{\mathbf{d}}_{k}} \right)\left[ j \right] \right|}^{2}} \right]\le 2l\sigma _{\omega }^{2}}{\mathop{\max }}\,I\left( F\left( {{\mathbf{d}}_{k}} \right)\left[ j \right]:{{\mathbf{y}}_{k}}\left[ j \right] \right).
\end{equation} 
For $l$ independent $d_{i} \in {\rm {\mathcal C}{\mathcal{N}}}\left(0,\sigma _{d_{i} }^{2} \right)$ subcarriers,
\begin{equation} \label{ZEqnNum500526} 
\begin{split}
   I\left( F\left( \mathbf{d} \right):\mathbf{y} \right)&={{H}_{\text{diff}}}\left( \mathbf{y} \right)-{{H}_{\text{diff}}}\left( \left. \mathbf{y} \right|F\left( \mathbf{d} \right) \right) \\ 
 & =\sum\nolimits_{l}{{{H}_{\text{diff}}}\left( {{y}_{i}} \right)-{{H}_{\text{diff}}}\left( \left. {{y}_{i}} \right|{{z}_{i}} \right)} \\ 
 & =\sum\nolimits_{l}{{{H}_{\text{diff}}}\left( {{y}_{i}} \right)-{{H}_{\text{diff}}}\left( F\left( {{\Delta }_{i}} \right) \right)} \\ 
 & =\sum\nolimits_{l}{{{\log }_{2}}\left( 1+\tfrac{\sigma _{{{\omega }_{i}}}^{2}{{\left| F\left( {{T}_{i}}\left( {{\mathcal{N}}_{i}} \right) \right) \right|}^{2}}}{\sigma _{\mathcal{N}}^{2}} \right)},  
\end{split}
\end{equation} 
where
\begin{equation} \label{74)} 
{{H}_{\text{diff}}}\left( F\left( \mathbf{d} \right),\mathbf{y} \right)={{H}_{\text{diff}}}\left( F\left( \mathbf{d} \right) \right)+{{H}_{\text{diff}}}\left( \left. \mathbf{y} \right|F\left( \mathbf{d} \right) \right)\le {{H}_{\text{diff}}}\left( F\left( \mathbf{d} \right) \right)+{{H}_{\text{diff}}}\left( \mathbf{y} \right).\end{equation} 
(Note: If the $l$ Gaussian subcarriers are not completely mutually independent, then $\le $ stands in the second line of Equation \eqref{ZEqnNum500526}, whereas if the $l$ subcarriers are derived from not the optimal zero-mean circular symmetric complex Gaussian random ${\rm {\mathcal C}{\mathcal{N}}}$ distribution, then $\le $ stands in the third line of Equation \eqref{ZEqnNum500526}.)

The input maximization leads to the following sum capacity in an AMQD-MQA setting:
\begin{equation} \label{75)} 
C_{{\rm sum}} \left({\rm {\mathcal N}}\right)=\mathop{\max }\limits_{\forall i} \sum _{l}\log _{2} \left(1+{\textstyle\frac{\sigma _{\omega _{i} }^{2} \left|F\left(T_{i} \left({\rm {\mathcal N}}_{i} \right)\right)\right|^{2} }{\sigma _{{\rm {\mathcal N}}}^{2} }} \right) ,                      
\end{equation} 
where ${\textstyle\frac{1}{l}} \sum _{l}\left|F\left(T_{i} \left({\rm {\mathcal N}}_{i} \right)\right)\right|^{2}  >{\textstyle\frac{1}{l}} \sum _{l}\left|T_{i} \left({\rm {\mathcal N}}_{i} \right)\right|^{2}  $ \cite{ref4}, with a total transmit variance constraint
\begin{equation} \label{76)} 
{\textstyle\frac{1}{n}} \sum _{n}\sigma _{\omega _{i} }^{2}  =\sigma _{\omega }^{2} =\sigma _{\omega _{0} }^{2} ,{\rm \; }\sigma _{\omega _{i} }^{2} \ge 0.                                
\end{equation} 
Hence, for the $l$ sub-channels with a constant nonzero modulation variance $\sigma _{\omega _{i} }^{2} >0$, one gets
\begin{equation} \label{77)} 
{\textstyle\frac{1}{l}} \sum _{l}\sigma _{\omega _{i} }^{2}  =\sigma _{\omega }^{2} <\sigma _{\omega _{0} }^{2} , 
\end{equation} 
where $\sigma _{\omega _{0} }^{2} $ is the modulation variance of $z_{i} $.

For two users, $U_{1} $ and $U_{2} $, the ${\rm{C}}$ capacity region $\left(R_{1} ,R_{2} \right)$ of AMQD-MQA is as follows:
\begin{equation} \label{78)} 
\begin{split}
  & {{R}_{1}}\le {{I}_{\text{MQA}}}\left( {{z}_{1}}:\left. y,F\left( T\left( \mathcal{N} \right) \right) \right|{{z}_{2}} \right)={{I}_{\text{MQA}}}\left( {{z}_{1}}:\left. y \right|F\left( T\left( \mathcal{N} \right) \right),{{z}_{2}} \right), \\ 
 & {{R}_{2}}\le {{I}_{\text{MQA}}}\left( {{z}_{2}}:\left. y,F\left( T\left( \mathcal{N} \right) \right) \right|{{z}_{1}} \right)={{I}_{\text{MQA}}}\left( {{z}_{2}}:\left. y \right|F\left( T\left( \mathcal{N} \right) \right),{{z}_{1}} \right), \\ 
\end{split}
\end{equation} 
and
\begin{equation} \label{79)} 
R_{1} +R_{2} \le I_{{\rm MQA}} \left(z_{1} ,z_{2} :y,F\left(T\left({\rm {\mathcal N}}\right)\right)\right)=I_{{\rm MQA}} \left(z_{1} ,z_{2} :\left. y\right|F\left(T\left({\rm {\mathcal N}}\right)\right)\right). 
\end{equation} 
Thus, for user $U_{k} $,
\begin{equation} \label{ZEqnNum838919} 
R_{k}^{{\rm MQA}} \le \mathop{\max }\limits_{\forall i} \sum _{l}\log _{2} \left(1+{\textstyle\frac{\sigma _{\omega _{i} }^{2} \left|F\left(T_{i} \left({\rm {\mathcal N}}_{i} \right)\right)\right|^{2} }{\sigma _{{\rm {\mathcal N}}}^{2} }} \right) ,{\rm \; }k\in 1,\ldots K,                 
\end{equation} 
the sum rate (the overall throughput rate of the users) $R_{{\rm sum}}^{{\rm MQA}} $ is calculated as,
\begin{equation} \label{ZEqnNum651242} 
R_{{\rm sum}}^{{\rm MQA}} =\sum _{K}R_{k}^{{\rm MQA}}  \le \mathop{\max }\limits_{\forall i} \sum _{l}\log _{2} \left(1+{\textstyle\frac{\sigma _{\omega _{i} }^{2} \left|F\left(T_{i} \left({\rm {\mathcal N}}_{i} \right)\right)\right|^{2} }{\sigma _{{\rm {\mathcal N}}}^{2} }} \right) , 
\end{equation} 
and the symmetric rate (the common rate at which all users can have a simultaneous reliable communication) $R_{{\rm sym}}^{{\rm MQA}} $ is calculated as
\begin{equation} \label{ZEqnNum243409} 
R_{{\rm sym}}^{{\rm MQA}} \le {\textstyle\frac{1}{K}} \mathop{\max }\limits_{\forall i} \sum _{l}\log _{2} \left(1+{\textstyle\frac{\sigma _{\omega _{i} }^{2} \left|F\left(T_{i} \left({\rm {\mathcal N}}_{i} \right)\right)\right|^{2} }{\sigma _{{\rm {\mathcal N}}}^{2} }} \right) .                             
\end{equation} 
For $K$ users $U_{1} ,\ldots U_{K} $, from the results of Equations \eqref{ZEqnNum838919} and \eqref{ZEqnNum651242} trivially follows $R_{{\rm k}}^{{\rm MQA}} $ and $R_{{\rm sym}}^{{\rm MQA}} $; hence, the sum capacity in AMQD-MQA is expressed as 
\begin{equation} \label{ZEqnNum269002} 
C_{{\rm sum}}^{{\rm MQA}} \left({\rm {\mathcal N}}\right)=\mathop{\max }\limits_{\left(R_{1}^{{\rm MQA}} ,\ldots ,R_{K}^{{\rm MQA}} \right)\in {\rm{C}}} \sum _{K}R_{k}^{{\rm MQA}}  =\mathop{\max }\limits_{\forall i} \sum _{l}\log _{2} \left(1+{\textstyle\frac{\sigma _{\omega _{i} }^{2} \left|F\left(T_{i} \left({\rm {\mathcal N}}_{i} \right)\right)\right|^{2} }{\sigma _{{\rm {\mathcal N}}}^{2} }} \right) , 
\end{equation} 
whereas the symmetric capacity (the maximum common rate at which all users can have a simultaneous reliable communication) is expressed as
\begin{equation} \label{ZEqnNum201124} 
C_{{\rm sym}}^{{\rm MQA}} \left({\rm {\mathcal N}}\right)=\mathop{\max }\limits_{\left(R_{{\rm sym}}^{{\rm MQA}} ,\ldots ,R_{{\rm sym}}^{{\rm MQA}} \right)\in {\rm{C}}} R_{{\rm sym}}^{{\rm MQA}} ={\textstyle\frac{1}{K}} \mathop{\max }\limits_{\forall i} \sum _{l}\log _{2} \left(1+{\textstyle\frac{\sigma _{\omega _{i} }^{2} \left|F\left(T_{i} \left({\rm {\mathcal N}}_{i} \right)\right)\right|^{2} }{\sigma _{{\rm {\mathcal N}}}^{2} }} \right) .           
\end{equation} 
The results on the ${\rm{C}}$ capacity region of $\left(R_{1} ,R_{2} \right)$ of $U_{1} $ and $U_{2} $ in AMQD-MQA are summarized in \fref{fig3}. The corner points $C_{1} $ and $C_{2} $ identify the maximal rates at with a single user can communicate. The line between the two corner points represents that trade-off between the rates of users $U_{1} $ and $U_{2} $, at which simultaneously reliable transmission is possible.

\begin{center}
\begin{figure*}[!h]
\begin{center}
\includegraphics[angle = 0,width=0.7\linewidth]{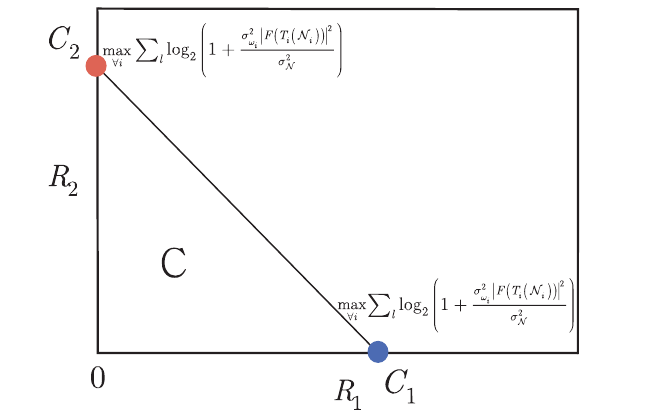}
\caption{The ${\rm{C}}$ capacity region of the AMQD-MQA with two users $U_{1} $ and $U_{2} $. The transmission is realized through $l$ subcarriers, each having a constant modulation variance $\sigma _{\omega }^{2} $ per quadrature components. The two users communicate over the Gaussian quantum channel with rates $R_{1} $ and $R_{2} $. At the corner points $C_{1} $ and $C_{2} $ (red and blue dots), only one user is allowed to transmit and all degrees of freedom is allocated to that user.} 
 \label{fig3}
 \end{center}
\end{figure*}
\end{center}

Assuming a $d$-dimensional input ${{\mathbf{z}}_{k}}\in \mathcal{C}\mathcal{N}\left( 0,{{\mathbf{K}}_{{{\mathbf{z}}_{k}}}} \right)$, and vector ${{\mathbf{d}}_{k}}\in \mathcal{C}\mathcal{N}\left( 0,{{\mathbf{K}}_{{{\mathbf{d}}_{k}}}} \right)$ for the subcarriers of each $U_{k} $, the capacity region ${\rm{C}}$ is expressed as follows:
\begin{equation} \label{ZEqnNum356274} 
{{R}_{k}}\le {{\log }_{2}}\det \left( {{\mathbf{I}}_{d}}+\tfrac{1}{2\sigma _{\mathcal{N}}^{2}}F\left( \mathbf{T}\left( \mathcal{N} \right) \right){{\mathbf{K}}_{{{\mathbf{d}}_{k}}}}F{{\left( \mathbf{T}\left( \mathcal{N} \right) \right)}^{\dagger }} \right),
\end{equation} 
and
\begin{equation} \label{ZEqnNum575993} 
{{R}_{1}}+{{R}_{2}}\le {{\log }_{2}}\det \left( {{\mathbf{I}}_{d}}+\tfrac{1}{2\sigma _{\mathcal{N}}^{2}}F\left( \mathbf{T}\left( \mathcal{N} \right) \right){{\mathbf{K}}_{{{\mathbf{d}}_{k}}}}F{{\left( \mathbf{T}\left( \mathcal{N} \right) \right)}^{\dagger }} \right),
\end{equation} 
where ${{\mathbf{I}}_{d}}$ is the $d$-dimensional identity matrix. Equations \eqref{ZEqnNum356274} and \eqref{ZEqnNum575993} are simultaneously maximized for the ${{\mathbf{z}}_{k}}\in \mathcal{C}\mathcal{N}\left( 0,{{\mathbf{K}}_{{{\mathbf{z}}_{k}}}} \right)$ Gaussian random vectors with different covariance matrices ${{\mathbf{K}}_{{{\mathbf{z}}_{k}}}}$. The capacity region ${\rm{C}}$ is determined by the ${{\mathbf{K}}_{{{\mathbf{z}}_{k}}}}$ covariance matrices of the users which are allowed to be determined by the users (or, equivalently by an encoder, depending on the actual setting of AMQD-MQA). 

These results conclude that in the AMQD-MQA CVQKD multiple access, the rates all of the $K$ users can simultaneously maximized, which concludes the proof.
\end{proof}

\subsection{Rates at Partial Channel Side Information}
\begin{theorem}
(Reliable simultaneous transmission at partial channel side information). The AMQD-MQA provides reliable simultaneous communication for the $K$ users at partial channel side information, with a sum rate $R_{{\rm sum}}^{{\rm MQA}} \le \mathop{\max }\limits_{\forall i} {\rm {\mathbb{E}}}\left[\sum _{l}\log _{2} \left(1+\sigma _{\omega _{i} }^{2} {\left|F\left(T_{i} \left({\rm {\mathcal N}}_{i} \right)\right)\right|^{2}  \mathord{\left/{\vphantom{\left|F\left(T_{i} \left({\rm {\mathcal N}}_{i} \right)\right)\right|^{2}  \sigma _{{\rm {\mathcal N}}}^{2} }}\right.\kern-\nulldelimiterspace} \sigma _{{\rm {\mathcal N}}}^{2} } \right) \right]$ and a symmetric rate $R_{sym}^{MQA} \le {\textstyle\frac{1}{K}} \mathop{\max }\limits_{\forall i} {\rm {\mathbb{E}}}\left[\sum _{l}\log _{2} \left(1+{\sigma _{\omega _{i} }^{2} \left|F\left(T_{i} \left({\rm {\mathcal N}}_{i} \right)\right)\right|^{2}  \mathord{\left/{\vphantom{\sigma _{\omega _{i} }^{2} \left|F\left(T_{i} \left({\rm {\mathcal N}}_{i} \right)\right)\right|^{2}  \sigma _{{\rm {\mathcal N}}}^{2} }}\right.\kern-\nulldelimiterspace} \sigma _{{\rm {\mathcal N}}}^{2} } \right) \right],$ where $T_{i} $ is a random complex variable.
\end{theorem}
\begin{proof}
The partial channel side information arises from the fact that the parties are not able to track exactly the $T_{i} $ transmittance coefficients of the ${\rm {\mathcal N}}_{i} $ sub-channels, only a close approximation is possible. However, because the CVQKD protocols are operating in the low-SNR regimes, this approximation also could be optimal for information transmission over the Gaussian sub-channels. In the allocation phase, only the $\lambda $ Lagrangian multiplier of the eavesdropper has to be taken into consideration that finally leads to $\nu _{{\text{Eve}}} ={1 \mathord{\left/{\vphantom{1 \lambda }}\right.\kern-\nulldelimiterspace} \lambda } $, the security parameter of the optimal Gaussian collective attack \cite{ref4}. As an important corollary, it is not a requirement for the parties to have an exact knowledge about states the sub-channels, that is, the $T_{i} $ transmittance coefficients. Furthermore, if Alice has no channel side information, in the low-SNR regimes, the optimal solution is to use a constant modulation variance for all Gaussian subcarriers. It is exactly the case in an AMQD setting because it is not a reasonable assumption a practical CVQKD scenario that the parties have full channel side information to allocate perfectly the modulation variances of the subcarriers. In the low-SNR regimes, the performance that can be achieved by a constant modulation variance is very close to the rates that can be obtained at an exact allocation. Here, we derive the transmission rates that can be reached if the parties have no exact knowledge about the sub-channels. The results trivially follow Theorem 1.

For the mutual information $I\left(z_{k} :y_{k} ,F\left(T\left({\rm {\mathcal N}}\right)\right)\right)$ and $I\left(z_{k} :F\left(T\left({\rm {\mathcal N}}\right)\right)\right)=0$, and from the chain rule of mutual information
\begin{equation} \label{87)} 
\begin{split}
   I\left( {{z}_{k}}:{{y}_{k}},F\left( T\left( \mathcal{N} \right) \right) \right)&=I\left( {{z}_{k}}:F\left( T\left( \mathcal{N} \right) \right) \right)+I\left( {{z}_{k}}:\left. y \right|F\left( T\left( \mathcal{N} \right) \right) \right) \\ 
 & =I\left( {{z}_{k}}:\left. y \right|F\left( T\left( \mathcal{N} \right) \right) \right).  
\end{split}
\end{equation} 
As one can immediately conclude, if Alice has no exact knowledge of the sub-channel transmittance coefficients, the ideal input distribution remains the zero-mean circular symmetric complex Gaussian random inputs $z_{k} =x_{k} +{\rm i}p_{k} \in {\rm {\mathcal C}{\mathcal{N}}}\left(0,{\rm {\mathbb{E}}}\left[\left|z_{k} \right|^{2} \right]\right)$, independently from the actual SNR values.

The mutual information can be maximized by the zero-mean ${\rm {\mathcal C}{\mathcal{N}}}$ distribution, which leads to 
\begin{equation} \label{88)} 
I\left(z_{k} :\left. y\right|F\left(T\left({\rm {\mathcal N}}\right)\right)\right)=\sum _{l}\log _{2} \left(1+{\textstyle\frac{\sigma _{\omega _{i} }^{2} \left|F\left(T_{i} \left({\rm {\mathcal N}}_{i} \right)\right)\right|^{2} }{\sigma _{{\rm {\mathcal N}}}^{2} }} \right) .                         
\end{equation} 
Assuming two users, $U_{1} $ and $U_{2} $, the rates of the users are
\begin{equation} \label{89)} 
\begin{split}
  & {{R}_{1}}\le \mathbb{E}\left[ I\left( {{z}_{1}}:\left. y \right|F\left( T\left( \mathcal{N} \right) \right),{{z}_{2}} \right) \right], \\ 
 & {{R}_{2}}\le \mathbb{E}\left[ I\left( {{z}_{2}}:\left. y \right|F\left( T\left( \mathcal{N} \right) \right),{{z}_{1}} \right) \right], \\ 
\end{split}
\end{equation} 
and
\begin{equation} \label{90)} 
R_{1} +R_{2} \le {\rm {\mathbb{E}}}\left[I\left(z_{1} ,z_{2} :\left. y\right|F\left(T\left({\rm {\mathcal N}}\right)\right)\right)\right].                                
\end{equation} 
The rate of a given user depends on the distribution of $z_{k} $, which for an AWGN channel with circular symmetric Gaussian random noise $\Delta _{i} \in {\rm {\mathcal C}{\mathcal{N}}}\left(0,\sigma _{\Delta _{i} }^{2} \right)$ picks up its maximum if the input $z_{k} $ is drawn from a ${\rm {\mathcal C}{\mathcal{N}}}\left(0,\sigma _{z_{k} }^{2} \right)$ random distribution (for optimality conditions, see Lemma 1). Assuming $d$ AMQD blocks for the optimization over all possible $x'_{k} $ distributions, the upper bound $R_{k} \le {\textstyle\frac{1}{d}} \mathop{\max }\limits_{\forall x'} I\left(x'_{k} :y\right)$ follows, and if $x'_{k} =z_{k} \in {\rm {\mathcal C}{\mathcal{N}}}\left(0,\sigma _{z_{k} }^{2} \right)$, then $R_{k} ={\textstyle\frac{1}{d}} \mathop{\max }\limits_{\forall x'} I\left(z_{k} :y\right)$. From the SNR of the $l$ Gaussian sub-channels with ${\rm SNR}={\sigma _{\omega _{i} }^{2}  \mathord{\left/{\vphantom{\sigma _{\omega _{i} }^{2}  \sigma _{{\rm {\mathcal N}}_{i} }^{2} }}\right.\kern-\nulldelimiterspace} \sigma _{{\rm {\mathcal N}}_{i} }^{2} } $ per sub-channel ${\rm {\mathcal N}}_{i} $, for $d$ AMQD blocks (each with $l$ subcarriers), and from the law of large numbers follows that 
\begin{equation} \label{91)} 
\begin{split}
   {{R}_{k}}&\le \tfrac{1}{d}\underset{\forall i}{\mathop{\max }}\,I\left( {{z}_{k}}:y \right) \\ 
 & =\underset{d\to \infty }{\mathop{\lim }}\,\tfrac{1}{d}\underset{\forall i}{\mathop{\max }}\,\sum\nolimits_{d}{{{\log }_{2}}}\left( 1+{{\left| F\left( {{T}_{k}}\left( \mathcal{N} \right) \right) \right|}^{2}}\text{SNR} \right) \\ 
 & =\underset{d\to \infty }{\mathop{\lim }}\,\tfrac{1}{d}\underset{\forall i}{\mathop{\max }}\,\sum\nolimits_{d}{\sum\nolimits_{l}{{{\log }_{2}}}}\left( 1+{{\left| {{F}_{k,i}}\left( {{T}_{k,i}}\left( {{\mathcal{N}}_{i}} \right) \right) \right|}^{2}}\text{SNR} \right) \\ 
 & =\underset{\forall i}{\mathop{\max }}\,\mathbb{E}\left[ \sum\nolimits_{l}{{{\log }_{2}}\left( 1+{{\left| F\left( {{T}_{i}}\left( {{\mathcal{N}}_{i}} \right) \right) \right|}^{2}}\text{SNR} \right)} \right] \\ 
 & =\underset{\forall i}{\mathop{\max }}\,\mathbb{E}\left[ \sum\nolimits_{l}{{{\log }_{2}}\left( 1+\tfrac{\sigma _{{{\omega }_{i}}}^{2}{{\left| F\left( {{T}_{i}}\left( {{\mathcal{N}}_{i}} \right) \right) \right|}^{2}}}{\sigma _{\mathcal{N}}^{2}} \right)} \right].  
\end{split}
\end{equation} 
These arguments yield that the transmission at a partial channel side information is characterized by exactly the same sum and symmetric rates as in the case of the full channel side information:
\begin{equation} \label{92)} 
R_{{\rm sum}}^{{\rm MQA}} =\sum _{K}R_{k}^{{\rm MQA}}  \le \mathop{\max }\limits_{\forall i} {\rm {\mathbb{E}}}\left[\sum _{l}\log _{2} \left(1+{\textstyle\frac{\sigma _{\omega _{i} }^{2} \left|F\left(T_{i} \left({\rm {\mathcal N}}_{i} \right)\right)\right|^{2} }{\sigma _{{\rm {\mathcal N}}}^{2} }} \right) \right], 
\end{equation} 
and to symmetric rate $R_{{\rm sym}}^{{\rm MQA}} $,
\begin{equation} \label{93)} 
R_{{\rm sym}}^{{\rm MQA}} \le {\textstyle\frac{1}{K}} \mathop{\max }\limits_{\forall i} {\rm {\mathbb{E}}}\left[\sum _{l}\log _{2} \left(1+{\textstyle\frac{\sigma _{\omega _{i} }^{2} \left|F\left(T_{i} \left({\rm {\mathcal N}}_{i} \right)\right)\right|^{2} }{\sigma _{{\rm {\mathcal N}}}^{2} }} \right) \right].                           
\end{equation} 
Hence, the corresponding capacities for the $U_{k} $ of this channel are as follows:
\begin{equation} \label{94)} 
\begin{split}
   C_{\text{sum}}^{\text{MQA}}\left( \mathcal{N} \right)&=\underset{\sigma _{F\left( {{\mathbf{d}}_{k}}\left[ j \right] \right)}^{2}=\mathbb{E}\left[ {{\left\| F\left( {{\mathbf{d}}_{k}} \right)\left[ j \right] \right\|}^{2}} \right]\le 2l\sigma _{\omega }^{2}}{\mathop{\max }}\,I\left( {{z}_{k}}:\left. y \right|F\left( T\left( \mathcal{N} \right) \right) \right) \\ 
 & =\underset{\forall i}{\mathop{\max }}\,\mathbb{E}\left[ \sum\nolimits_{l}{{{\log }_{2}}\left( 1+\tfrac{\sigma _{{{\omega }_{i}}}^{2}{{\left| F\left( {{T}_{i}}\left( {{\mathcal{N}}_{i}} \right) \right) \right|}^{2}}}{\sigma _{\mathcal{N}}^{2}} \right)} \right],  
\end{split}
\end{equation} 
and
\begin{equation} \label{95)} 
\begin{split}
   C_{\text{sym}}^{\text{MQA}}\left( \mathcal{N} \right)&=\tfrac{1}{K}\underset{\sigma _{F\left( {{\mathbf{d}}_{k}}\left[ j \right] \right)}^{2}=\mathbb{E}\left[ {{\left\| F\left( {{\mathbf{d}}_{k}}\left[ j \right] \right) \right\|}^{2}} \right]\le 2l\sigma _{\omega }^{2}}{\mathop{\max }}\,I\left( {{z}_{k}}:\left. y \right|F\left( T\left( \mathcal{N} \right) \right) \right) \\ 
 & =\tfrac{1}{K}\underset{\forall i}{\mathop{\max }}\,\mathbb{E}\left[ \sum\nolimits_{l}{{{\log }_{2}}\left( 1+\tfrac{\sigma _{{{\omega }_{i}}}^{2}{{\left| F\left( {{T}_{i}}\left( {{\mathcal{N}}_{i}} \right) \right) \right|}^{2}}}{\sigma _{\mathcal{N}}^{2}} \right)} \right],  
\end{split}
\end{equation} 
which confirms that the rates of the users are exactly the same as that of Theorem 1. It is a particularly convenient practical benefit that arises directly from the AMQD modulation. 

For the $K\to K$ scenario in terms of $K$-dimensional vectors $\mathbf{z},\mathbf{y}$, the situation is as follows. The channel output can be rewritten as
\begin{equation} \label{96)} 
\begin{split}
   \mathbf{y}&=F\left( \mathbf{T}\left( \mathcal{N} \right) \right)\mathbf{z}+F\left( \Delta  \right) \\ 
 & =F\left( \mathbf{T}\left( \mathcal{N} \right) \right)F\left( {{F}^{-1}}\left( \mathbf{z} \right) \right)+F\left( \Delta  \right) \\ 
 & =F\left( \mathbf{T}\left( \mathcal{N} \right) \right)F\left( \mathbf{d} \right)+F\left( \Delta  \right),  
\end{split}
\end{equation} 
where 
\begin{equation} \label{97)} 
F\left( \mathbf{T}\left( \mathcal{N} \right) \right)={{\left[ F\left( {{T}_{1}}\left( \mathcal{N} \right) \right),\ldots ,F\left( {{T}_{l}}\left( \mathcal{N} \right) \right) \right]}^{T}},
\end{equation} 
\begin{equation} \label{98)} 
\mathbf{z}\left( \mathcal{N} \right)={{\left[ {{z}_{1}},\ldots ,{{z}_{K}} \right]}^{T}}\in \mathcal{C}\mathcal{N}\left( 0,{{\mathbf{K}}_{\mathbf{z}}} \right),
\end{equation} 
\begin{equation} \label{99)} 
F\left( \Delta  \right)={{\left[ F\left( {{\Delta }_{1}} \right),\ldots ,F\left( {{\Delta }_{l}} \right) \right]}^{T}}\in \mathcal{C}\mathcal{N}\left( 0,{{\mathbf{K}}_{F\left( \Delta  \right)}} \right)=\mathcal{C}\mathcal{N}\left( 0,2\sigma _{\mathcal{N}}^{2}{{\mathbf{I}}_{K}} \right),
\end{equation} 
and
\begin{equation} \label{100)} 
F\left( \mathbf{d} \right)={{\left[ F\left( {{d}_{1}} \right),\ldots ,F\left( {{d}_{l}} \right) \right]}^{T}}\in \mathcal{C}\mathcal{N}\left( 0,{{\mathbf{K}}_{F\left( \mathbf{d} \right)}} \right).
\end{equation} 
From these, the mutual information is
\begin{equation} \label{101)} 
\begin{split}
   I\left( \mathbf{z}:\left. \mathbf{y} \right|F\left( \mathbf{T}\left( \mathcal{N} \right) \right) \right)&={{H}_{\text{diff}}}\left( \mathbf{y} \right)-{{H}_{\text{diff}}}\left( \left. \mathbf{y} \right|\mathbf{z} \right) \\ 
 & ={{H}_{\text{diff}}}\left( \mathbf{y} \right)-{{H}_{\text{diff}}}\left( F\left( \Delta  \right) \right) \\ 
 & ={{H}_{\text{diff}}}\left( \mathbf{y} \right)-K{{\log }_{2}}\left( \pi e2\sigma _{{{\mathcal{N}}_{i}}}^{2} \right).  
\end{split}
\end{equation} 
The $K$-dimensional zero-mean circular symmetric complex Gaussian random $\mathbf{d}\in \mathcal{C}\mathcal{N}\left( 0,{{\mathbf{K}}_{\mathbf{d}}} \right)$ input vectors preserve the entropy maximizer property \cite{ref17,ref18,ref19} (more precisely, it maximizes the $H_{{\rm diff}} $ differential entropy function among all $K$-dimensional complex, arbitrarily distributed random vectors $\mathbf{d}$ that have a given covariance matrix ${{\mathbf{K}}_{\mathbf{d}}}$); hence, ${{H}_{\text{diff}}}\left( \mathbf{d} \right)={{\log }_{2}}\left( \det \left( \pi e{{\mathbf{K}}_{\mathbf{d}}} \right) \right)$. The output has the covariance matrix ${{\mathbf{K}}_{\mathbf{y}}}=F\left( \mathbf{T}\left( \mathcal{N} \right) \right){{\mathbf{K}}_{\mathbf{d}}}F{{\left( \mathbf{T}\left( \mathcal{N} \right) \right)}^{\dagger }}+2\sigma _{{{\mathcal{N}}_{i}}}^{2}{{\mathbf{I}}_{K}}$. 

This leads to the mutual information,
\begin{equation} \label{ZEqnNum153278}
\begin{split}
   I\left( \mathbf{z}:\left. \mathbf{y} \right|F\left( \mathbf{T}\left( \mathcal{N} \right) \right) \right)&={{\log }_{2}}\left( {{\left( \pi e \right)}^{K}} \right)\det \left( F\left( \mathbf{T}\left( \mathcal{N} \right) \right){{\mathbf{K}}_{\mathbf{d}}}F{{\left( \mathbf{T}\left( \mathcal{N} \right) \right)}^{\dagger }}+2\sigma _{{{\mathcal{N}}_{i}}}^{2}{{\mathbf{I}}_{K}} \right) \\ 
 & \text{ }-K{{\log }_{2}}\left( \pi e2\sigma _{{{\mathcal{N}}_{i}}}^{2} \right) \\ 
 & ={{\log }_{2}}\det \left( {{\mathbf{I}}_{K}}+\tfrac{F\left( \mathbf{T}\left( \mathcal{N} \right) \right){{\mathbf{K}}_{\mathbf{d}}}F{{\left( \mathbf{T}\left( \mathcal{N} \right) \right)}^{\dagger }}}{2\sigma _{\mathcal{N}}^{2}} \right).  
\end{split}
\end{equation}

The equality in the first line of Equation \eqref{ZEqnNum153278} follows from the fact that the subcarrier vector $\mathbf{d}$ has the distribution $\mathcal{C}\mathcal{N}\left( 0,{{\mathbf{K}}_{\mathbf{d}}} \right)$, which immediately reveals that the encoder, in fact, does not need to know in an exact form the channel transmittance vector $\mathbf{T}\left( \mathcal{N} \right)$ because the ${\rm {\mathcal C}{\mathcal{N}}}$ input distribution leads to the optimal rate. It yields a $K$-dimensional $\mathbf{z}\in \mathcal{C}\mathcal{N}\left( 0,{{\mathbf{K}}_{\mathbf{z}}} \right)$:
\begin{equation} \label{103)} 
C=\underset{{{\mathbf{K}}_{\mathbf{d}}}:Tr\left( {{\mathbf{K}}_{\mathbf{d}}} \right)\le 2l\sigma _{\omega }^{2}}{\mathop{\max }}\,\mathbb{E}\left[ {{\log }_{2}}\det \left( {{\mathbf{I}}_{K}}+\tfrac{F\left( \mathbf{T}\left( \mathcal{N} \right) \right){{\mathbf{K}}_{\mathbf{d}}}F{{\left( \mathbf{T}\left( \mathcal{N} \right) \right)}^{\dagger }}}{2\sigma _{\mathcal{N}}^{2}} \right) \right],
\end{equation} 
which finally concludes the proof.
\end{proof}

\subsection{Optimality}
\begin{lemma}
(Optimality of AMQD-MQA). The AMQD-MQA simultaneously maximizes the transmission rates of all $K$ users over the Gaussian quantum channel.
\end{lemma}
\begin{proof}
The Gaussian quantum channel ${\rm {\mathcal N}}$ can be modeled as a complex channel, with complex noise $\Delta \in {\rm {\mathcal C}{\mathcal{N}}}\left(0,\sigma _{\Delta }^{2} \right)$ in the phase space ${\rm {\mathcal S}}$. For a given sub-channel ${\rm {\mathcal N}}_{i} $, the extension follows with noise $\Delta _{i} \in {\rm {\mathcal C}{\mathcal{N}}}\left(0,\sigma _{\Delta _{i} }^{2} \right)$. Let us evaluate the $H_{{\rm diff}} \left(\cdot \right)$ differential entropy (see Equation \eqref{ZEqnNum203317}) of the noise variable $\Delta _{i} =\Delta _{x,i} +\Delta _{p,i} $, where $\Delta _{x,i} \in {\rm {\mathcal{N}}}\left(0,\sigma _{{\rm {\mathcal N}}_{i} }^{2} \right)$ and $\Delta _{p,i} \in {\rm {\mathcal{N}}}\left(0,\sigma _{{\rm {\mathcal N}}_{i} }^{2} \right)$ are i.i.d. zero-mean Gaussian random noise on the position ($x$) and momentum ($p$) quadratures of the Gaussian sub-channel ${\rm {\mathcal N}}_{i} $, respectively. The function $H_{{\rm diff}} \left(\Delta _{i} \right)$ is as follows:
\begin{equation} \label{104)}
\begin{split}
   {{H}_{\text{diff}}}\left( {{\Delta }_{i}} \right)&={{H}_{\text{diff}}}\left( \operatorname{\text{Re}}\left( {{\Delta }_{i}} \right) \right)+{{H}_{\text{diff}}}\left( \operatorname{\text{Im}}\left( {{\Delta }_{i}} \right) \right) \\ 
 & ={{H}_{\text{diff}}}\left( {{\Delta }_{x,i}} \right)+{{H}_{\text{diff}}}\left( {{\Delta }_{p,i}} \right) \\ 
 & ={{\log }_{2}}\left( \pi e\sigma _{{{\Delta }_{i}}}^{2} \right) \\ 
 & ={{\log }_{2}}\left( \pi e\mathbb{E}\left[ {{\left| {{\Delta }_{i}} \right|}^{2}} \right] \right) \\ 
 & ={{\log }_{2}}\left( \pi e2\sigma _{{{\mathcal{N}}_{i}}}^{2} \right).  
\end{split}
\end{equation}

For the subcarrier variable $d_{i} \in {\rm {\mathcal C}{\mathcal{N}}}\left(0,\sigma _{d_{i} }^{2} \right),$ the differential entropy is evaluated as
\begin{equation} \label{105)} 
\begin{split}
   {{H}_{\text{diff}}}\left( {{d}_{i}} \right)&={{H}_{\text{diff}}}\left( \operatorname{\text{Re}}\left( {{d}_{i}} \right) \right)+{{H}_{\text{diff}}}\left( \operatorname{\text{Im}}\left( {{d}_{i}} \right) \right) \\ 
 & ={{H}_{\text{diff}}}\left( {{x}_{{{d}_{i}}}} \right)+{{H}_{\text{diff}}}\left( {{p}_{{{d}_{i}}}} \right) \\ 
 & ={{\log }_{2}}\left( \pi e\sigma _{{{d}_{i}}}^{2} \right) \\ 
 & ={{\log }_{2}}\left( \pi e\mathbb{E}\left[ {{\left| {{d}_{i}} \right|}^{2}} \right] \right) \\ 
 & ={{\log }_{2}}\left( \pi e2\sigma _{{{\omega }_{i}}}^{2} \right).  
\end{split}
\end{equation} 
To prove the optimality of AMQD-MQA, we use the fact that for the subcarrier variable $d$, the density function $F_{d} \left(\cdot \right)$ picks up the values on a support set $S$, whereas for $u\notin S$, $F_{d} \left(u\right)=0$ \cite{ref17}. The $d_{i} $ variable has maximal differential entropy on $S$ among all possible $x'$ probability distributions with second moment condition $2\sigma _{\omega }^{2} $, only if the following equation holds:
\begin{equation} \label{ZEqnNum668658} 
\int\limits_{S}f \left(u\right)F_{d} \left(u\right)du=2\sigma _{\omega }^{2} .                                       
\end{equation} 
Taking the second moment condition with respect to the i.i.d. Gaussian random quadrature components $x_{d} ,p_{d} $, one obtains
\begin{equation} \label{107)} 
\int\limits_{S}f \left(u\right)F_{x_{d} } \left(u\right)du=\sigma _{\omega }^{2} ,                                      
\end{equation} 
and
\begin{equation} \label{108)} 
\int\limits_{S}f \left(u\right)F_{p_{d} } \left(u\right)du=\sigma _{\omega }^{2} .                                       
\end{equation} 
In particular, the subcarrier variable $d$ is entropy maximizer only if the density function $F_{d} \left(\cdot \right)$ is in perfect coincidence with the Gaussian probability density function. To see it, let the $F_{d} $ density of variable $d$ be given in the following formula:
\begin{equation} \label{ZEqnNum929066} 
F_{d} =e^{\left(\gamma _{0} -1+\sum _{2}\gamma _{i} f_{i} \left(u\right) \right)} ,\, u\in S.                                   
\end{equation} 
By choosing the density function to the density of the ${\rm {\mathcal C}{\mathcal{N}}}$ distribution, that is, for an $d\in {\rm {\mathcal C}{\mathcal{N}}}\left(0,\sigma _{d}^{2} \right)$ variable, Equation \eqref{ZEqnNum929066} can be rewritten as
\begin{equation} \label{110)} 
F_{d} \left(d\right)={\textstyle\frac{1}{2\pi \sigma _{\omega }^{2} }} e^{{\textstyle\frac{-\left(\left|d\right|^{2} \right)}{2\sigma _{\omega }^{2} }} } =f\left(x_{d} ,p_{d} \right)={\textstyle\frac{1}{2\pi \sigma _{\omega }^{2} }} e^{{\textstyle\frac{-\left(x_{d} ^{2} +p_{d} ^{2} \right)}{2\sigma _{\omega }^{2} }} } , 
\end{equation} 
which leads to those $\gamma _{0} ,\gamma _{1} ,\gamma _{2} $, such that Equation \eqref{ZEqnNum668658} is satisfied. For any other distribution, the moment condition of Equation \eqref{ZEqnNum668658} cannot be met; thus, variable $d$ has maximal entropy only if it is a zero-mean circular symmetric complex Gaussian random variable with variance $2\sigma _{\omega _{i} }^{2} $. This clearly demonstrates that the $i$-th subcarrier $d_{i} $ maximizes the entropy only if it is drawn from a ${\rm {\mathcal C}{\mathcal{N}}}\left(0,\sigma _{d}^{2} \right)$ distribution, with i.i.d. zero-mean Gaussian random quadrature components. The results trivially follow for $\mathbf{d}\in \mathcal{C}\mathcal{N}\left( 0,{{\mathbf{K}}_{d}} \right)$, ${{\mathbf{K}}_{d}}=\mathbb{E}\left[ \mathbf{d}{{\mathbf{d}}^{\dagger }} \right]$, because the entropy is maximized by the structure of the ${\rm {\mathcal C}{\mathcal{N}}}$ distribution, which finally leads to the differential entropy ${{H}_{\text{diff}}}\left( \mathbf{d} \right)={{\log }_{2}}\det \left( \pi e{{\mathbf{K}}_{d}} \right)$, which precisely coincidences with the theoretical entropy maximum. 

Because all users in AMQD-MQA transmit the information via ${\rm {\mathcal C}{\mathcal{N}}}\left(0,\sigma _{d}^{2} \right)$-distributed subcarriers with i.i.d. Gaussian random quadrature components $x_{d_{i} } ,p_{d_{i} } \in {\rm {\mathcal{N}}}\left(0,\sigma _{\omega _{i} }^{2} \right)$, the optimality of the scheme is immediately concluded.
\end{proof}

\section{Compensation of a Nonideal Gaussian Modulation}
\label{sec4}
\begin{theorem}
(Compensation of a nonideal Gaussian input modulation). A nonideal Gaussian input modulation can be compensated by $\nu _{i} +\nu _{\min } \left(1-{\rm {\mathcal G}}\left(\delta \right)_{p\left(x\right)} \right)$ sub-channel coefficients, where $\nu _{i} ={\sigma _{{\rm {\mathcal N}}}^{2}  \mathord{\left/{\vphantom{\sigma _{{\rm {\mathcal N}}}^{2}  \left|F\left(T_{i} \left({\rm {\mathcal N}}_{i} \right)\right)\right|^{2} }}\right.\kern-\nulldelimiterspace} \left|F\left(T_{i} \left({\rm {\mathcal N}}_{i} \right)\right)\right|^{2} } $ is the coefficient of ${\rm {\mathcal N}}_{i} ,$ $\nu _{\min } =\min \left\{\nu _{1} ,\ldots ,\nu _{l} \right\}$, ${\rm {\mathcal G}}\left(\delta \right)_{p\left(x\right)} <1$, and $0\le \delta \le 1$. 
\end{theorem}
\begin{proof}
The proof is organized as follows. First we assume that the modulation variance can be varied. Then we reveal that for a constant modulation variance the problem is analogous to the lifting up the $\nu _{i} $ levels of the sub-channels. 

Let $d_{i} $ be the ideal input variable of ${\left| \phi _{i}  \right\rangle} \in {\rm {\mathcal S}}$ drawn from a ${\rm {\mathcal C}{\mathcal{N}}}\left(0,2\sigma _{\omega }^{2} \right)$ distribution, and let $d'$ be the model variable of the noisy output Gaussian subcarrier $\left| {{\phi '_{i}}} \right\rangle \in \mathcal{S}$. Let the quantity $\xi ,{\rm \; }0\le \xi \le 1$ be the mean square estimation error of the input $d_{i} $ from output ${{d}'_{i}}$ at modulation variance $2\sigma _{\omega }^{2} $ and in presence of Gaussian noise $\Delta _{i} $, presented as
\begin{equation} \label{111)} 
\begin{split}
   \xi \left( {{{{d}'_{i}}}} \right)&=\mathbb{E}\left[ {{\left| {{d}_{i}}-f\left( {{{{d}'_{i}}}},2\sigma _{\omega }^{2} \right) \right|}^{2}} \right] \\ 
 & =\mathbb{E}\left[ {{\left| {{d}_{i}}-f\left( \sqrt{2\sigma _{\omega }^{2}}{{d}_{i}}+{{\Delta }_{i}},2\sigma _{\omega }^{2} \right) \right|}^{2}} \right],  
\end{split}
\end{equation} 
where $d_{i} $ is characterized with a unit modulation variance and
\begin{equation} \label{112)} 
f\left(\sqrt{2\sigma _{\omega }^{2} } d_{i} +\Delta _{i} ,2\sigma _{\omega }^{2} \right)={\rm {\mathbb{E}}}\left[\left. d_{i} \right|\sqrt{2\sigma _{\omega }^{2} } d_{i} +\Delta _{i} \right].                         
\end{equation} 
Particularly, for an ideal Gaussian-modulated subcarrier CV $d_{i} $, $\xi \left(\cdot \right)$ is defined as
\begin{equation} \label{ZEqnNum290060} 
\xi \left(d_{i} \right)={\textstyle\frac{1}{1+2\sigma _{\omega }^{2} }} , 
\end{equation} 
and the $\xi ^{-1} \left(\cdot \right)$ inverse function of $\xi \left(\cdot \right)$ for the ideal Gaussian modulation is evaluated as
\begin{equation} \label{ZEqnNum463708} 
\xi ^{-1} \left(\mho \right)={\textstyle\frac{1-\mho }{\mho }} , 
\end{equation} 
where $\mho >0$. The $\sigma _{\omega }^{2} $ constant modulation variance of the subcarrier quadratures for an arbitrary distribution $p\left(x\right)$ can be rewritten as follows:
\begin{equation} \label{ZEqnNum120013} 
\sigma _{\omega }^{2} =\nu _{{\text{Eve}}} -\nu _{\min } {\rm {\mathcal G}}\left(\delta \right)_{p\left(x\right)} ,                                     
\end{equation} 
where variable $\delta \ge 0$ quantifies the $p\left(x\right)$ input distribution, as $0\le \delta \le 1$ for an arbitrary $p\left(x\right)$ distribution, whereas  $\delta $ is arbitrary for an ideal Gaussian-modulated input \cite{ref22}, $\nu _{{\text{Eve}}} $ is the security parameter that identifies an optimal Gaussian collective attack \cite{ref4}, \cite{ref12,ref13}, calculated as
\begin{equation} \label{ZEqnNum791007} 
\nu _{{\text{Eve}}} ={\textstyle\frac{1}{\lambda }} ,                                                    
\end{equation} 
where $\lambda $ is the Lagrange multiplier \cite{ref4}, \cite{ref17}, calculated as
\begin{equation} \label{117)} 
\lambda =\left|F\left(T_{{\rm {\mathcal N}}}^{{\rm *}} \right)\right|^{2} ={\textstyle\frac{1}{n}} \sum _{i=0}^{n-1}\left|\sum _{k=0}^{n-1}T_{k}^{{\rm *}} e^{{\textstyle\frac{-{\rm i}2\pi ik}{n}} }  \right|^{2}  ,        
\end{equation} 
where $T_{{\rm {\mathcal N}}}^{{\rm *}} $ is the expected transmittance of the $n$ sub-channels (i.e., all sub-channels are taken into consideration) under an optimal Gaussian collective attack. 

According to the definition of AMQD modulation \cite{ref4}, from $\lambda $, and the $\sigma _{\omega _{i} }^{2} $ modulation variances of the ${\rm {\mathcal N}}_{i} $ sub-channels, a Lagrangian can be constructed as
\begin{equation} \label{118)} 
{\rm {\mathcal L}}\left(\lambda ,\sigma _{\omega _{1} }^{2} \ldots \sigma _{\omega _{n} }^{2} \right)=\sum _{i=1}^{n}\log _{2}  \left(1+{\textstyle\frac{\sigma _{\omega _{i} }^{2} \left|F\left(T_{i} \left({\rm {\mathcal N}}_{i} \right)\right)\right|^{2} }{\sigma _{{\rm {\mathcal N}}}^{2} }} \right)-\lambda \sum _{i=1}^{n}\sigma _{\omega _{i} }^{2}  .                
\end{equation} 
Using the Kuhn-Tucker condition \cite{ref17,ref18,ref19}, it follows that ${\partial {\rm {\mathcal L}} \mathord{\left/{\vphantom{\partial {\rm {\mathcal L}} \partial \sigma _{\omega _{i} }^{2} }}\right.\kern-\nulldelimiterspace} \partial \sigma _{\omega _{i} }^{2} } =0$ only if the $i$-th sub-channel gets a nonzero modulation variance, $\sigma _{\omega _{i} }^{2} >0$, whereas ${\partial {\rm {\mathcal L}} \mathord{\left/{\vphantom{\partial {\rm {\mathcal L}} \partial \sigma _{\omega _{i} }^{2} }}\right.\kern-\nulldelimiterspace} \partial \sigma _{\omega _{i} }^{2} } \le 0$ if the sub-channel gets zero modulation variance, $\sigma _{\omega _{i} }^{2} =0$ \cite{ref20,ref21,ref22}. 

After some calculations, the average modulation variance leads to
\begin{equation} \label{119)} 
\sigma _{\omega }^{2} ={\textstyle\frac{1}{n}} \sum _{i=1}^{n}\left(\nu _{{\text{Eve}}} -{\textstyle\frac{\sigma _{{\rm {\mathcal N}}}^{2} }{\left|F\left(T_{i} \left({\rm {\mathcal N}}_{i} \right)\right)\right|^{2} }} \right) ={\textstyle\frac{1}{n}} \sum _{i=1}^{n}\left(\nu _{{\text{Eve}}} -\nu _{i} \right) .                   
\end{equation} 
As it can be verified, the optimal solution for this problem is a constant modulation variance $\sigma _{\omega }^{2} $ for those $l$ ${\rm {\mathcal N}}_{i} $ sub-channels, for which $\nu _{i} <\nu _{{\text{Eve}}} $ is satisfied:
\begin{equation} \label{120)} 
\sigma _{\omega }^{2} ={\textstyle\frac{1}{l}} \sum _{i=1}^{l}\left(\nu _{{\text{Eve}}} -{\textstyle\frac{\sigma _{{\rm {\mathcal N}}}^{2} }{\mathop{\max }\limits_{i} \left|F\left(T_{i} \left({\rm {\mathcal N}}_{i} \right)\right)\right|^{2} }} \right) =\nu _{{\text{Eve}}} -\min \left(\nu _{i} \right)=\nu _{{\text{Eve}}} -\nu _{\min } .   
\end{equation} 
In other words, $\nu _{\min } $ in Equation \eqref{ZEqnNum120013} identifies the $\min \left\{\nu _{1} ,\ldots ,\nu _{l} \right\}$ minimum of the $\nu _{i} $ sub-channel coefficients, where
\begin{equation} \label{121)} 
\nu _{i} ={\sigma _{{\rm {\mathcal N}}}^{2}  \mathord{\left/{\vphantom{\sigma _{{\rm {\mathcal N}}}^{2}  \left|F\left(T_{i} \left({\rm {\mathcal N}}_{i} \right)\right)\right|^{2} }}\right.\kern-\nulldelimiterspace} \left|F\left(T_{i} \left({\rm {\mathcal N}}_{i} \right)\right)\right|^{2} } . 
\end{equation} 
The function ${\rm {\mathcal G}}\left(\delta \right)_{p\left(x\right)} $ in Equation \eqref{ZEqnNum120013} quantifies the deviation of the input modulation (input distribution $p\left(x\right)$) from the ideal Gaussian random distribution in terms of mean square estimation error \cite{ref22}, precisely as:
\begin{equation} \label{122)} 
{\rm {\mathcal G}}\left(\delta \right)_{p\left(x\right)} ={\textstyle\frac{1}{\nu _{\min } \kappa }} -\xi ^{-1} \left(\delta \right)_{p\left(x\right)} ,                                  
\end{equation} 
where $0<\kappa <{\textstyle\frac{1}{\nu _{\min } }} $.

For an ideal Gaussian input modulation,
\begin{equation} \label{123)} 
\sigma _{\omega }^{2} +\nu _{\min } {\rm {\mathcal G}}^{{\rm *}} \left(\delta \right)_{p\left(x\right)} ={\textstyle\frac{1}{\kappa }} ,                                       
\end{equation} 
where ${\rm {\mathcal G}}^{{\rm *}} \left(\delta \right)_{p\left(x\right)} $ is expressed as
\begin{equation} \label{124)} 
\begin{split}
   {{\mathcal{G}}^{\text{*}}}{{\left( \delta  \right)}_{p\left( x \right)}}&=\tfrac{1}{{{\nu }_{\min }}\kappa }-{{\xi }^{\text{*}}}^{-1}{{\left( \delta  \right)}_{p\left( x \right)}} \\ 
 & =\tfrac{1}{{{\nu }_{\min }}\kappa }-\tfrac{1-{{\nu }_{\min }}\kappa }{{{\nu }_{\min }}\kappa } \\ 
 & =1,  
\end{split}
\end{equation} 
where
\begin{equation} \label{125)} 
{{\xi }^{\text{*}}}^{-1}{{\left( \delta  \right)}_{p\left( x \right)}}=\tfrac{1-{{\nu }_{\min }}\kappa }{{{\nu }_{\min }}\kappa }
\end{equation} 
for an ideal Gaussian input \cite{ref17,ref18,ref19}, \cite{ref22}, see Equation \eqref{ZEqnNum463708}. Thus, for an ideal Gaussian modulation,
\begin{equation} \label{126)} 
\sigma _{\omega }^{2} =\nu _{{\text{Eve}}} -\nu _{\min } , \sigma _{\kappa }^{2} =0, 
\end{equation} 
and 
\begin{equation} \label{127)} 
\kappa ={\textstyle\frac{1}{\sigma _{\omega }^{2} +\nu _{\min } }} .                                              
\end{equation} 
Hence,
\begin{equation} \label{128)} 
\begin{split}
   \tfrac{1}{\kappa }&=\sigma _{\omega }^{2}+{{\nu }_{\min }} \\ 
 & =\left( {{\nu }_{\text{Eve}}}-{{\nu }_{\min }}{{\mathcal{G}}^{\text{*}}}{{\left( \delta  \right)}_{p\left( x \right)}} \right)+{{\nu }_{\min }} \\ 
 & ={{\nu }_{\text{Eve}}}.  
\end{split}
\end{equation} 
Assuming that the modulation variance is not pre-determined, an additional modulation variance $\sigma _{\kappa }^{2} $ that is required for the compensation of a nonideal Gaussian modulation (for a nonideal input modulation with distribution $p\left(x\right)$, i.e., assuming a truncated modulation that approximates the ideal Gaussian modulation by a discretization), is as follows:
\begin{equation} \label{129)} 
\begin{split}
   \tfrac{1}{\kappa }&=\left( {{\nu }_{\text{Eve}}}-{{\nu }_{\min }}\mathcal{G}{{\left( \delta  \right)}_{p\left( x \right)}} \right)+{{\nu }_{\min }} \\ 
 & ={{\nu }_{\text{Eve}}}-{{\nu }_{\min }}\left( \mathcal{G}{{\left( \delta  \right)}_{p\left( x \right)}}-1 \right) \\ 
 & ={{\nu }_{\text{Eve}}}+{{\nu }_{\min }}\left( 1-\mathcal{G}{{\left( \delta  \right)}_{p\left( x \right)}} \right) \\ 
 & =\left( \sigma _{\omega }^{2}+\sigma _{\kappa }^{2} \right)+{{\nu }_{\min }} \\ 
 & ={{\nu }_{\text{Eve}}}+\sigma _{\kappa }^{2},  
\end{split}
\end{equation} 
where $\kappa ={\textstyle\frac{1}{\nu _{{\text{Eve}}} +\nu _{\min } \left(1-{\rm {\mathcal G}}\left(\delta \right)_{p\left(x\right)} \right)}} $, $\sigma _{\kappa }^{2} =\nu _{\min } \left(1-{\rm {\mathcal G}}\left(\delta \right)_{p\left(x\right)} \right)$ and ${\rm {\mathcal G}}\left(\delta \right)_{p\left(x\right)} $ is evaluated as
\begin{equation} \label{130)} 
{\rm {\mathcal G}}\left(\delta \right)_{p\left(x\right)} ={\textstyle\frac{1}{\nu _{\min } \kappa }} -\xi ^{-1} \left(\delta \right)_{p\left(x\right)} {\rm <}<{\rm {\mathcal G}}^{{\rm *}} \left(\delta \right)_{p\left(x\right)} , 
\end{equation} 
where ${\rm {\mathcal G}}\left(\delta \right)_{p\left(x\right)} <1$ and
\begin{equation} \label{ZEqnNum302076} 
{{\xi }^{-1}}{{\left( \delta  \right)}_{p\left( x \right)}}>{{\xi }^{\text{*}}}^{-1}{{\left( \delta  \right)}_{p\left( x \right)}}.
\end{equation} 
The actual value of $\xi ^{-1} \left(\delta \right)_{p\left(x\right)} $ in Equation \eqref{ZEqnNum302076} depends on the probability distribution $p\left(x\right)$.

Since the AMQD-MQA scheme uses fixed modulation variance $\sigma _{\omega }^{2} $ for all sub-channels, the coefficient $\nu _{i} $ of Gaussian sub-channel ${\rm {\mathcal N}}_{i} $ has to be lifted up in the calculations by a nonzero $\nu _{\kappa } >0$ additional term, precisely
\begin{equation} \label{132)} 
\begin{split}
   {{\nu }_{\kappa }}&={{\nu }_{\min }}\left( 1-\mathcal{G}{{\left( \delta  \right)}_{p\left( x \right)}} \right) \\ 
 & ={{\nu }_{\min }}\left( 1-\left( \tfrac{1}{{{\nu }_{\min }}\kappa }-{{\xi }^{-1}}{{\left( \delta  \right)}_{p\left( x \right)}} \right) \right).  
\end{split}
\end{equation}
Putting the pieces together, for a nonideal Gaussian input modulation, for each ${\rm {\mathcal N}}_{i} ,{\rm \; }i=1,\ldots ,l$, the $\nu _{i} $ coefficients are evaluated as
\begin{equation} \label{133)} 
\nu _{i} +\nu _{\min } \left(1-{\rm {\mathcal G}}\left(\delta \right)_{p\left(x\right)} \right)=\nu _{i} +\sigma _{\kappa }^{2} =\nu _{i} +\nu _{\kappa } .                         
\end{equation}

The steps of the compensation of a nonideal Gaussian input modulation is given in Algorithm 1. 

 \setcounter{algocf}{0}
\begin{algo}
  \DontPrintSemicolon
\caption{\textit{Compensation of a nonideal Gaussian input modulation}}
\begin{enumerate}
\item Determine $\nu _{\min } =\min \left\{\nu _{1} ,\ldots ,\nu _{l} \right\}$ of the $l$ Gaussian sub-channels, where $\nu _{i} ={\sigma _{{\rm {\mathcal N}}}^{2}  \mathord{\left/{\vphantom{\sigma _{{\rm {\mathcal N}}}^{2}  \left|F\left(T_{i} \left({\rm {\mathcal N}}_{i} \right)\right)\right|^{2} }}\right.\kern-\nulldelimiterspace} \left|F\left(T_{i} \left({\rm {\mathcal N}}_{i} \right)\right)\right|^{2} } $, $\nu _{i} <\nu _{{\text{Eve}}} $. From $\nu _{{\text{Eve}}} ={\textstyle\frac{1}{\lambda }} $ and $\nu _{\min } $, compute the constant modulation variance $\sigma _{\omega }^{2} $ of the Gaussian subcarriers as $\sigma _{\omega }^{2}={{\nu }_{\text{Eve}}}-{{\nu }_{\min }}{{\mathcal{G}}^{\text{*}}}{{\left( \delta  \right)}_{p\left( x \right)}}$, where ${{\mathcal{G}}^{\text{*}}}{{\left( \delta  \right)}_{p\left( x \right)}}=\tfrac{1}{{{\nu }_{\min }}\kappa }-{{\xi }^{\text{*}}}^{-1}{{\left( \delta  \right)}_{p\left( x \right)}}=1$, $\kappa ={\textstyle\frac{1}{\sigma _{\omega }^{2} +\nu _{\min } }} $, and $\delta $ is arbitrary. 

\item Let $p\left(x\right)$ be the distribution of the nonideal input. Choose $\kappa $ such that $\kappa ={\textstyle\frac{1}{\nu _{{\text{Eve}}} -\nu _{\min } \left({\rm {\mathcal G}}\left(\delta \right)_{p\left(x\right)} -1\right)}} $, where ${\rm {\mathcal G}}\left(\delta \right)_{p\left(x\right)} ={\textstyle\frac{1}{\nu _{\min } \kappa }} -\xi ^{-1} \left(\delta \right)_{p\left(x\right)} ,$ $0\le \delta \le 1$, ${{\xi }^{-1}}{{\left( \delta  \right)}_{p\left( x \right)}}>{{\xi }^{\text{*}}}^{-1}{{\left( \delta  \right)}_{p\left( x \right)}}$, and ${\rm {\mathcal G}}\left(\delta \right)_{p\left(x\right)} <1$.

\item Determine $\sigma _{\kappa }^{2} >0$ such that ${\textstyle\frac{1}{\kappa }} =\nu _{{\text{Eve}}} +\sigma _{\kappa }^{2} $.

\item Compute $\nu _{\kappa } =\nu _{\min } \left(1-{\rm {\mathcal G}}\left(\delta \right)_{p\left(x\right)} \right)$ and $\nu _{i} +\nu _{\kappa } $ for the Gaussian sub-channels ${\rm {\mathcal N}}_{i} ,{\rm \; }i=1,\ldots ,l$. 

\end{enumerate}
\end{algo}

The proof is concluded here.
\end{proof}

\begin{center}
\begin{figure*}[!h]
\begin{center}
\includegraphics[angle = 0,width=0.7\linewidth]{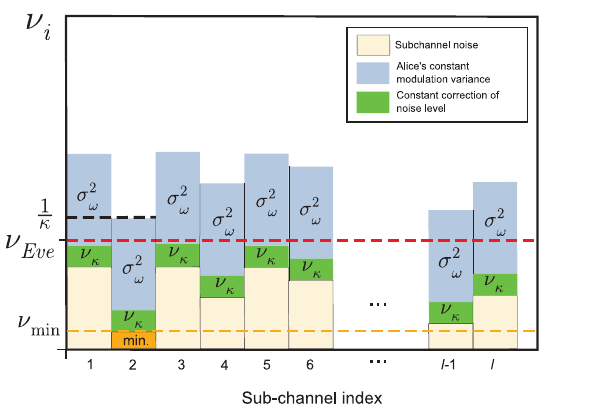}
\caption{The compensation of a nonideal Gaussian modulation. The supplemental constant $\nu _{\kappa } >0$ term is added to $\nu _{i} $ of the Gaussian sub-channels, $\nu _{i} +\nu _{\kappa } <\nu _{Eve} $, to overwhelm the modulation imperfections.} 
 \label{fig4}
 \end{center}
\end{figure*}
\end{center}

The compensation of a nonideal Gaussian modulation is depicted in \fref{fig4}. Depending on the difference $\kappa $ of the input modulation and the ideal Gaussian distribution ${\rm {\mathcal C}{\mathcal{N}}}\left(0,2\sigma _{\omega }^{2} \right)$, the noise level $\nu _{i} $ of each Gaussian sub-channels is lifted up to $\nu _{i} +\nu _{\min } \left(1-{\rm {\mathcal G}}\left(\delta \right)_{p\left(x\right)} \right)=\nu _{i} +\nu _{\kappa } $, for ${\rm {\mathcal N}}_{i} ,{\rm \; }i=1,\ldots ,l$. For an ideal Gaussian modulation, $\nu _{\kappa } =0$. 

The algorithm of the optimal constant modulation variance adaption within the AMQD modulation can be found in detail in \cite{ref4}, as we do not include the details here.

\section{Opportunistic Gaussian Modulation}
\label{sec5}
\begin{theorem}
(Diversity amplification via opportunistic Gaussian modulation for improved SNR). Let $\mathop{\max }\limits_{\forall i} \left|F\left(T_{k} \left({\rm {\mathcal N}}_{U_{k} } \right)\right)\right|^{2} =\mathop{\max }\limits_{\forall i} \sum _{s}\left|F\left(T_{k,i} \left({\rm {\mathcal N}}_{i} \right)\right)\right|^{2}  $ of ${\rm {\mathcal N}}_{U_{k} } =\left[{\rm {\mathcal N}}_{1} ,\ldots ,{\rm {\mathcal N}}_{s} \right]^{T} $ for user $U_{k} $, where $s\le l$ is the number of subcarriers of transmit user $U_{k} $ in the $j$-th AMQD block. By an opportunistic Gaussian modulation, the maximized transmittance coefficient of ${\rm {\mathcal N}}_{U_{k} } \left[j\right]$ can be increased to $\mathop{\max }\limits_{\forall i} \left|F\left(T'_{k} \left({\rm {\mathcal N}}_{U_{k} } \right)\right)\right|^{2} =\mathop{\max }\limits_{\forall i} \sum _{s}\left|\sqrt{a_{k,i} } e^{{\rm i}\theta _{k,i} } F\left(T_{k,i} \left({\rm {\mathcal N}}_{i} \right)\right)\right|^{2}  $, where $\sqrt{a_{k,i} } e^{{\rm i}\theta _{k,i} } $ is a complex variable, $a_{k,i} \ge 0$, $0\le \theta _{k,i} \le 2\pi $, and $\sum _{s}a_{k,i}  >1$. 
\end{theorem}
\begin{proof}
The opportunistic Gaussian modulation exploits a natural property of the Gaussian sub-channels, which arises from the diversity of the $T_{i} \left({\rm {\mathcal N}}_{i} \right)$ transmittance coefficients. The main idea behind opportunistic Gaussian modulation in AMQD-MQA is as follows: enhance the SNR of the Gaussian sub-channel ${\rm {\mathcal N}}_{i} $ by the amplification of $\left|T_{i} \left({\rm {\mathcal N}}_{i} \right)\right|^{2} $, if the distribution of $T_{i} \left({\rm {\mathcal N}}_{i} \right)$ allows it. Precisely, the opportunistic Gaussian modulation performs as a diversity amplification, which induces a much stronger randomization into the transmittance coefficients of the Gaussian sub-channels. It is particularly convenient because it leads to improved SNR for the logical sub-channel ${\rm {\mathcal N}}_{U_{k} } \left[j\right]=\left[{\rm {\mathcal N}}_{1} ,\ldots ,{\rm {\mathcal N}}_{s} \right]^{T} $ of user $U_{k} $ (where ${\rm {\mathcal N}}_{U_{k} } \left[j\right]$ stands for the set of ${\rm {\mathcal N}}_{i} $ Gaussian sub-channels that transmit the $s$ subcarriers of user $U_{k} $ in the $j$-th AMQD block). We propose the technique for setting 1 of AMQD-MQA (see \fref{fig1}).

Assume two users, $U_{1} $ and $U_{2} $, with rates $R_{1} \approx R_{2} $ and with squared magnitudes of the Fourier-transformed transmittance coefficients, as follows:
\begin{equation} \label{ZEqnNum572267} 
\left|F\left(T_{1} \left({\rm {\mathcal N}}_{U_{1} } \right)\right)\right|^{2} \approx \left|F\left(T_{2} \left({\rm {\mathcal N}}_{U_{2} } \right)\right)\right|^{2} .                              
\end{equation} 
In this case, the initial $\left|T_{i} \left({\rm {\mathcal N}}_{U_{i} } \right)\right|$ coefficients follow a stationary distribution; hence, the transmittance coefficients are centered around an average
\begin{equation} \label{ZEqnNum848051} 
c={\textstyle\frac{1}{K}} \sum _{k=1}^{K}{\textstyle\frac{1}{d}} \sum _{j=1}^{d}\left|F\left(T_{k} \left({\rm {\mathcal N}}_{U_{k} } \right)\left[j\right]\right)\right|  .                               
\end{equation} 
Let 
\begin{equation} \label{136)} 
O_{k,i} \left[j\right]=\sqrt{a_{k,i} } e^{{\rm i}\theta _{k,i} } , 
\end{equation} 
a complex variable that characterizes the opportunistic modulation \cite{ref17} in the $j$-th AMQD block, where 
\begin{equation} \label{137)} 
a_{k,i} \ge 0, 0\le \theta _{k,i} \le 2\pi , 
\end{equation} 
and $\sum _{s}a_{k,i}  >1$, for all transmit users $U_{k} $ of the AMQD block. 

Particular, for the $j$-th AMQD block, these parameters formulate an $s$-dimensional vector for user $U_{k} $:
\begin{equation} \label{138)} 
{{\mathbf{O}}_{k}}\left[ j \right]={{\left[ {{O}_{k,1}},\ldots ,{{O}_{k,s}} \right]}^{T}}={{\left[ \sqrt{{{a}_{k,1}}}{{e}^{i{{\theta }_{k,1}}}},\ldots ,\sqrt{{{a}_{k,s}}}{{e}^{i{{\theta }_{k,s}}}} \right]}^{T}}.
\end{equation} 
The AMQD-MQA opportunistic modulation is performed on the $z_{k,i} $ elements of $z_{k} $ of $U_{k} $:
\begin{equation} \label{ZEqnNum290170} 
\begin{split}
  & {{{\mathbf{{z}'_{k}}}}}\left[ j \right]={{\mathbf{O}}_{k}}{{\mathbf{z}}_{k}}\left[ j \right] \\ 
 & =\sqrt{{{a}_{k,i}}}{{e}^{i{{\theta }_{k,i}}}}{{z}_{k,i}},\text{ }i=0,\ldots s-1,  
\end{split}
\end{equation} 
where $s$ is the number of Gaussian subcarriers of $U_{k} $ in the $j$-th AMQD block.

The $j$-th block of ${{\mathbf{{y}'_{k}}}}$ of $U_{k} $ is evaluated as
\begin{equation} \label{140)} 
\begin{split}
   {{{\mathbf{{y}'_{k}}}}}\left[ j \right]&=F\left( {{T}_{k}}\left( {{\mathcal{N}}_{{{U}_{k}}}} \right) \right){{{\mathbf{{z}'_{k}}}}}\left[ j \right]+F\left( \Delta  \right)\left[ j \right] \\ 
 & =F\left( {{T}_{k}}\left( {{\mathcal{N}}_{{{U}_{k}}}} \right) \right){{\mathbf{O}}_{k}}{{\mathbf{z}}_{k}}\left[ j \right]+F\left( \Delta  \right)\left[ j \right] \\ 
 & =F\left( {{T}_{k}}\left( {{\mathcal{N}}_{{{U}_{k}}}} \right) \right)\left( \sum\nolimits_{s}{\sqrt{{{a}_{k,i}}}}{{e}^{i{{\theta }_{i}}}}{{z}_{k,i}} \right)+F\left( \Delta  \right)\left[ j \right] \\ 
 & =\left( \sum\nolimits_{s}{\sqrt{{{a}_{k,i}}}}{{e}^{i{{\theta }_{i}}}}F\left( {{T}_{k,i}}\left( {{\mathcal{N}}_{i}} \right) \right) \right){{z}_{k,i}}+F\left( \Delta  \right)\left[ j \right].  
\end{split}
\end{equation} 
In particular, the squared magnitude of the resulting transmittance coefficients of ${\rm {\mathcal N}}_{U_{k} } $ is 
\begin{equation} \label{141)} 
\left|F\left(T'_{k} \left({\rm {\mathcal N}}_{U_{k} } \right)\right)\right|^{2} =\sum _{s}\left|\sqrt{a_{k,i} } e^{{\rm i}\theta _{k,i} } F\left(T_{k,i} \left({\rm {\mathcal N}}_{i} \right)\right)\right|^{2}  ,                  
\end{equation} 
from which it immediately follows that the rate of user $U_{k} $ over ${\rm {\mathcal N}}_{U_{k} } $ (averaged over an AMQD block) is
\begin{equation} \label{ZEqnNum588805} 
R\left({\rm {\mathcal N}}_{U_{k} } \right)\le \mathop{\max }\limits_{\forall i} \sum _{s}\log _{2} \left(1+{\textstyle\frac{\sigma _{\omega _{i} }^{2} \left|\sqrt{a_{k,i} } e^{{\rm i}\theta _{k,i} } F\left(T_{k,i} \left({\rm {\mathcal N}}_{i} \right)\right)\right|^{2} }{\sigma _{{\rm {\mathcal N}}}^{2} }} \right) . 
\end{equation} 
Hence, for the transmittance coefficients $T_{k} \left({\rm {\mathcal N}}_{U_{k} } \right)$ and ${{T}_{k}}^{\prime }\left( {{\mathcal{N}}_{{{U}_{k}}}} \right)$, the relation
\begin{equation} \label{ZEqnNum545912} 
\begin{matrix}
  \underset{\forall i}{\mathop{\max }}\,\sum\nolimits_{s}{{{\left| \sqrt{{{a}_{k,i}}}{{e}^{i{{\theta }_{k,i}}}}F\left( {{T}_{k,i}}\left( {{\mathcal{N}}_{i}} \right) \right) \right|}^{2}}}>\underset{\forall i}{\mathop{\max }}\,\sum\nolimits_{s}{{{\left| F\left( {{T}_{k,i}}\left( {{\mathcal{N}}_{i}} \right) \right) \right|}^{2}}} \\ 
  \underset{\forall i}{\mathop{\max }}\,{{\left| F\left( {{{{T}'}}_{k}}\left( {{\mathcal{N}}_{{{U}_{k}}}} \right) \right) \right|}^{2}}>\underset{\forall i}{\mathop{\max }}\,{{\left| F\left( {{T}_{k}}\left( {{\mathcal{N}}_{{{U}_{k}}}} \right) \right) \right|}^{2}}, \\ 
\end{matrix}
\end{equation} 
follows; thus, at a given modulation variance $\sigma _{\omega _{i} }^{2} $, $\mathop{\max }\limits_{\forall i} R\left({\rm {\mathcal N}}_{U_{k} } \right)_{T'_{k} \left({\rm {\mathcal N}}_{U_{k} } \right)} >\mathop{\max }\limits_{\forall i} R\left({\rm {\mathcal N}}_{U_{k} } \right)_{T_{k} \left({\rm {\mathcal N}}_{U_{k} } \right)} $, that is,
\begin{equation} \label{144)} 
\mathop{\max }\limits_{\forall i} \sum _{s}\log _{2} \left(1+{\textstyle\frac{\sigma _{\omega _{i} }^{2} \left|\sqrt{a_{k,i} } e^{{\rm i}\theta _{k,i} } F\left(T_{k,i} \left({\rm {\mathcal N}}_{i} \right)\right)\right|^{2} }{\sigma _{{\rm {\mathcal N}}}^{2} }} \right) >\mathop{\max }\limits_{\forall i} \sum _{s}\log _{2} \left(1+{\textstyle\frac{\sigma _{\omega _{i} }^{2} \left|F\left(T_{k,i} \left({\rm {\mathcal N}}_{i} \right)\right)\right|^{2} }{\sigma _{{\rm {\mathcal N}}}^{2} }} \right) . 
\end{equation} 
The effect of opportunistic Gaussian modulation assuming a $d$-dimensional AMQD block code is illustrated in \fref{fig5}. The initial distribution of the $\left|F\left(T_{k} \left({\rm {\mathcal N}}_{U_{k} } \right)\right)\right|$ coefficients is close to stationary, around an average c, see Equations \eqref{ZEqnNum572267} and \eqref{ZEqnNum848051} , thus the diversity among the initial $\left|F\left(T_{k} \left({\rm {\mathcal N}}_{U_{k} } \right)\right)\right|$ transmittance coefficients is low. As a corollary of diversity amplification (see \eqref{ZEqnNum290170}), the distribution range of the transmittance coefficients is increased, which induces a strong trade-off between the transmittance coefficients of ${\rm {\mathcal N}}_{U_{k} } $, that results overall in higher SNR for $U_{k} $.

\begin{center}
\begin{figure*}[!h]
\begin{center}
\includegraphics[angle = 0,width=1\linewidth]{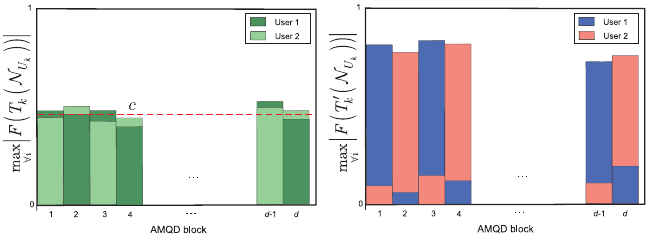}
\caption{Diversity amplification by opportunistic Gaussian modulation of the maximized transmittance coefficients of ${\rm {\mathcal N}}_{U_{k} } $, for an $d$-dimensional AMQD block code. (a) The distribution of the maximized transmittance coefficient is close to stationary around an average c. (b) The distribution of the $F\left(T_{k,i} \left({\rm {\mathcal N}}_{i} \right)\right)$ variables is optimized by opportunistic Gaussian modulation, which amplifies the diversity of the sub-channel transmittance coefficients. The stationary distribution of channel transmittance coefficients is randomized, which results in an optimized dynamic distribution range. The sub-channel diversity can be exploited, which leads to an improvement in the SNR.} 
 \label{fig5}
 \end{center}
\end{figure*}
\end{center}

The $\left\{a_{k,i} \ge 0\right\}$ and $\left\{\theta _{k,i} \right\}\in \left[0,2\pi \right]$ parameters can be dynamically adapted by the users so that the diversity amplification can be significant because the stationary (more precisely, close to a stationary) distribution of the $F\left(T_{k,i} \left({\rm {\mathcal N}}_{i} \right)\right)$ transmittance coefficients can be randomized. This randomization induces improvement in the overall channel transmittance $\left|F\left(T'_{k} \left({\rm {\mathcal N}}\right)\right)\right|$, which results in significantly better sub-channel diversity that can be exploited in a multiuser scenario. 

As an important corollary of the optimized dynamic distribution range, higher SNRs can be reached for ${\rm {\mathcal N}}_{U_{k} } $, which is particularly convenient in experimental CVQKD, which operates in the low-SNR regimes. Specifically, for the maximized SNR of ${\rm {\mathcal N}}_{U_{k} } $,
\begin{equation} \label{145)} 
\mathop{\max }\limits_{\forall i} \left|\sqrt{a_{k,i} } e^{{\rm i}\theta _{k,i} } F\left(T_{k,i} \left({\rm {\mathcal N}}_{i} \right)\right)\right|^{2} \cdot {\rm SNR}>\mathop{\max }\limits_{\forall i} \left|F\left(T_{k,i} \left({\rm {\mathcal N}}_{i} \right)\right)\right|^{2} \cdot {\rm SNR},             
\end{equation} 
where ${\rm SNR}={\sigma _{\omega _{i} }^{2}  \mathord{\left/{\vphantom{\sigma _{\omega _{i} }^{2}  \sigma _{{\rm {\mathcal N}}}^{2} }}\right.\kern-\nulldelimiterspace} \sigma _{{\rm {\mathcal N}}}^{2} } $.
\end{proof}

\section{Conclusions}
\label{sec6}
In contrast to DV QKD protocols, the CVQKD systems can be implemented within the current technological framework, which allows to perform unconditional secure communication over the already established optical communication networks by standard telecommunication and optical devices. In this work, we provided the AMQD-MQA multiple access scheme for CVQKD. The MQA is achieved by the AMQD modulation framework, which granulates the transmit information of the users into Gaussian subcarrier CV states. The rate allocation of the users is performed through the sophisticated handling of the subcarrier CVs and continuous unitary operations. We showed that in the AMQD-MQA, the users can optimally perform simultaneously reliable capacity-achieving communication over the Gaussian sub-channels. We developed an algorithm for the compensation of nonideal Gaussian modulation, which attenuates the imperfections of the input distribution. Finally, we investigated the diversity amplification for the Gaussian sub-channels, which improves the SNR of the users by opportunistic Gaussian modulation. The AMQD-MQA allows optimal multiple input--multiple output, capacity-achieving simultaneous transmission for the users, which is particularly convenient in an experimental long-distance CVQKD scenario, specifically in the crucial low-SNR regimes.

\section*{Acknowledgements}
This work was partially supported by the National Research Development and Innovation Office of Hungary (Project No. 2017-1.2.1-NKP-2017-00001), by the Hungarian Scientific Research Fund - OTKA K-112125 and in part by the BME Artificial Intelligence FIKP grant of EMMI (BME FIKP-MI/SC).

\bibliography{report}
\bibliographystyle{unsrt}

\newpage
\appendix
\setcounter{table}{0}
\setcounter{figure}{0}
\setcounter{equation}{0}
\setcounter{algocf}{0}
\renewcommand{\thetable}{\Alph{section}.\arabic{table}}
\renewcommand{\thefigure}{\Alph{section}.\arabic{figure}}
\renewcommand{\theequation}{\Alph{section}.\arabic{equation}}
\renewcommand{\thealgocf}{\Alph{section}.\arabic{algocf}}

\setlength{\arrayrulewidth}{0.1mm}
\setlength{\tabcolsep}{5pt}
\renewcommand{\arraystretch}{1.5}
\section{Appendix}
\subsection{Abbreviations}
\begin{description}
\item[AMQD] Adaptive Multicarrier Quadrature Division
\item[AWGN] Additive White Gaussian Noise
\item[BS] Beam Splitter
\item[CV] Continuous-Variable
\item[CVQFT] Continuous-Variable Quantum Fourier Transform
\item[DV] Discrete-Variable
\item[FFT] Fast Fourier Transform
\item[IFFT] Inverse Fast Fourier Transform
\item[MQA] Multiuser Quadrature Allocation
\item[OFDM] Orthogonal Frequency-Division Multiplexing
\item[OFDMA] OFDM Multiple Access
\item[SNR] Signal-to-Noise Ratio
\end{description}

\subsection{Notations}
The notations of the manuscript are summarized in \tref{tab2}.
\begin{center}
\begin{longtable}{||l|p{4.5in}||}
\caption{Summary of notations.}
\label{tab2}
\endfirsthead
\endhead
\hline
\textit{Notation} & \textit{Description} \\ \hline
$z\in {\rm {\mathcal C}{\mathcal{N}}}\left(0,\sigma _{z}^{2} \right)$ & The variable of a single-carrier Gaussian  CV state, ${\left| \varphi _{i}  \right\rangle} \in {\rm {\mathcal S}}$. Zero-mean, circular symmetric complex Gaussian random variable, $\sigma _{z}^{2} ={\rm {\mathbb{E}}}\left[\left|z\right|^{2} \right]=2\sigma _{\omega _{0} }^{2} $, with i.i.d. zero mean, Gaussian random quadrature components $x,p\in {\rm {\mathcal{N}}}\left(0,\sigma _{\omega _{0} }^{2} \right)$, where $\sigma _{\omega _{0} }^{2} $ is the variance.  \\ \hline 
$\Delta \in {\rm {\mathcal C}{\mathcal{N}}}\left(0,\sigma _{\Delta }^{2} \right)$ & The noise variable of the Gaussian channel ${\rm {\mathcal N}}$, with i.i.d. zero-mean, Gaussian random noise components on the position and momentum quadratures $\Delta _{x} ,\Delta _{p} \in {\rm {\mathcal{N}}}\left(0,\sigma _{{\rm {\mathcal N}}}^{2} \right)$, $\sigma _{\Delta }^{2} ={\rm {\mathbb{E}}}\left[\left|\Delta \right|^{2} \right]=2\sigma _{{\rm {\mathcal N}}}^{2} $. \\ \hline 
$d\in {\rm {\mathcal C}{\mathcal{N}}}\left(0,\sigma _{d}^{2} \right)$ & The variable of a Gaussian subcarrier CV state, ${\left| \phi _{i}  \right\rangle} \in {\rm {\mathcal S}}$. Zero-mean, circular symmetric Gaussian random variable, $\sigma _{d}^{2} ={\rm {\mathbb{E}}}\left[\left|d\right|^{2} \right]=2\sigma _{\omega }^{2} $, with i.i.d. zero mean, Gaussian random quadrature components $x_{d} ,p_{d} \in {\rm {\mathcal{N}}}\left(0,\sigma _{\omega }^{2} \right)$, where $\sigma _{\omega }^{2} $ is the modulation variance of the Gaussian subcarrier CV state.  \\ \hline 
$F^{-1} \left(\cdot \right)={\text{CVQFT}}^{\dag } \left(\cdot \right)$ & The inverse CVQFT transformation, applied by the encoder, continuous-variable unitary operation. \\ \hline 
$F\left(\cdot \right)={\text{CVQFT}}\left(\cdot \right)$ & The CVQFT transformation, applied by the decoder, continuous-variable unitary operation. \\ \hline 
$F^{-1} \left(\cdot \right)={\rm IFFT}\left(\cdot \right)$ & Inverse FFT transform, applied by the encoder. \\ \hline 
$\sigma _{\omega _{0} }^{2} $ & Single-carrier modulation variance. \\ \hline 
$\sigma _{\omega }^{2} ={\textstyle\frac{1}{l}} \sum _{l}\sigma _{\omega _{i} }^{2}  $ & Multicarrier modulation variance. Average modulation variance of the $l$ Gaussian sub-channels ${\rm {\mathcal N}}_{i} $.  \\ \hline 
$| {{\phi }_{i}} \rangle $ & The $i$-th Gaussian subcarrier CV of user $U_{k} $, $\left| {{\phi }_{i}} \right\rangle =\left| \text{IFFT}\left( {{z}_{k,i}} \right) \right\rangle =\left| {{F}^{-1}}\left( {{z}_{k,i}} \right) \right\rangle =\left| {{d}_{i}} \right\rangle $, where IFFT is the inverse fast Fourier transform, ${\left| \phi _{i}  \right\rangle} \in {\rm {\mathcal S}}$, $d_{i} \in {\rm {\mathcal C}{\mathcal{N}}}\left(0,\sigma _{d_{i} }^{2} \right)$, $\sigma _{d_{i} }^{2} ={\rm {\mathbb{E}}}\left[\left|d_{i} \right|^{2} \right]$, $d_{i} =x_{d_{i} } +{\rm i}p_{d_{i} } $, $x_{d_{i} } \in {\rm {\mathcal{N}}}\left(0,\sigma _{\omega _{F} }^{2} \right)$, $p_{d_{i} } \in {\rm {\mathcal{N}}}\left(0,\sigma _{\omega _{F} }^{2} \right)$ are i.i.d. zero-mean Gaussian random quadrature components, and $\sigma _{\omega _{F} }^{2} $ is the variance of the Fourier transformed Gaussian state. \\ \hline 
${\left| \varphi _{k,i}  \right\rangle}$ & The decoded single-carrier CV of user $U_{k} $ from the subcarrier CV, ${\left| \varphi _{k,i}  \right\rangle} ={\text{CVQFT}}\left({\left| \phi _{i}  \right\rangle} \right)$ also expressed as $F\left({\left| d_{i}  \right\rangle} \right)={\left| F\left(F^{-1} \left(z_{k,i} \right)\right) \right\rangle} ={\left| z_{k,i}  \right\rangle} .$ \\ \hline 
${\rm {\mathcal N}}$ & Gaussian quantum channel. \\ \hline 
${\rm {\mathcal N}}_{i} ,i=1,\ldots ,n$ & Gaussian sub-channels. \\ \hline 
$T\left({\rm {\mathcal N}}\right)$ & Channel transmittance, normalized complex random variable, $T\left({\rm {\mathcal N}}\right)=\text{Re}T\left({\rm {\mathcal N}}\right)+{\rm i}\text{Im}T\left({\rm {\mathcal N}}\right)\in {\rm {\mathcal C}}$. The real part identifies the position quadrature transmission, the imaginary part identifies the transmittance of the position quadrature. \\ \hline 
$T_{i} \left({\rm {\mathcal N}}_{i} \right)$ & Transmittance coefficient of Gaussian sub-channel ${\rm {\mathcal N}}_{i} $, $T_{i} \left({\rm {\mathcal N}}_{i} \right)=\text{Re}\left(T_{i} \left({\rm {\mathcal N}}_{i} \right)\right)+{\rm i}\text{Im}\left(T_{i} \left({\rm {\mathcal N}}_{i} \right)\right)\in {\rm {\mathcal C}}$, quantifies the position and momentum quadrature transmission, with (normalized) real and imaginary parts $0\le \text{Re}T_{i} \left({\rm {\mathcal N}}_{i} \right)\le {1 \mathord{\left/{\vphantom{1 \sqrt{2} }}\right.\kern-\nulldelimiterspace} \sqrt{2} } ,$ $0\le \text{Im}T_{i} \left({\rm {\mathcal N}}_{i} \right)\le {1 \mathord{\left/{\vphantom{1 \sqrt{2} }}\right.\kern-\nulldelimiterspace} \sqrt{2} } $, where $\text{Re}T_{i} \left({\rm {\mathcal N}}_{i} \right)=\text{Im}T_{i} \left({\rm {\mathcal N}}_{i} \right)$.  \\ \hline 
$T_{Eve} $ & Eve's transmittance, $T_{Eve} =1-T\left({\rm {\mathcal N}}\right)$. \\ \hline 
$T_{Eve,i} $ & Eve's transmittance for the $i$-th subcarrier CV. \\ \hline 
${\rm {\mathcal A}}\subseteq K$ & The subset of allocated users, ${\rm {\mathcal A}}\subseteq K$. Only the allocated users can transmit information in a given (particularly the $j$-th) AMQD block. The cardinality of subset ${\rm {\mathcal A}}$ is $\left|{\rm {\mathcal A}}\right|$. \\ \hline 
$U_{k} ,{\rm \; }k=1,\ldots ,\left|{\rm {\mathcal A}}\right|$ & An allocated user from subset ${\rm {\mathcal A}}\subseteq K$. \\ \hline 
$\mathbf{z}$ & A $d$-dimensional, zero-mean, circular symmetric complex random Gaussian vector, $\mathbf{z}=\mathbf{x}+i\mathbf{p}={{\left( {{z}_{1}},\ldots ,{{z}_{d}} \right)}^{T}}$, that models $d$ Gaussian CV input states, $\mathcal{C}\mathcal{N}\left( 0,{{\mathbf{K}}_{\mathbf{z}}} \right)$, ${{\mathbf{K}}_{\mathbf{z}}}=\mathbb{E}\left[ \mathbf{z}{{\mathbf{z}}^{\dagger }} \right]$, where $z_{i} =x_{i} +{\rm i}p_{i} $, $\mathbf{x}={{\left( {{x}_{1}},\ldots ,{{x}_{d}} \right)}^{T}}$, $\mathbf{p}={{\left( {{p}_{1}},\ldots ,{{p}_{d}} \right)}^{T}}$, with $x_{i} \in {\rm {\mathcal{N}}}\left(0,\sigma _{\omega _{0} }^{2} \right)$, $p_{i} \in {\rm {\mathcal{N}}}\left(0,\sigma _{\omega _{0} }^{2} \right)$ i.i.d. zero-mean Gaussian random variables. \\ \hline 
$\mathbf{d}={{F}^{-1}}\left( \mathbf{z} \right)$ & An $l$-dimensional, zero-mean, circular symmetric complex random Gaussian vector of the $l$ Gaussian subcarrier CVs, $\mathcal{C}\mathcal{N}\left( 0,{{\mathbf{K}}_{\mathbf{d}}} \right)$, ${{\mathbf{K}}_{\mathbf{d}}}=\mathbb{E}\left[ \mathbf{d}{{\mathbf{d}}^{\dagger }} \right]$, $\mathbf{d}={{\left( {{d}_{1}},\ldots ,{{d}_{l}} \right)}^{T}}$, $d_{i} =x_{i} +{\rm i}p_{i} $, $x_{i} ,p_{i} \in {\rm {\mathcal{N}}}\left(0,\sigma _{\omega _{F} }^{2} \right)$ are i.i.d. zero-mean Gaussian random variables, $\sigma _{\omega _{F} }^{2} ={1 \mathord{\left/{\vphantom{1 \sigma _{\omega _{0} }^{2} }}\right.\kern-\nulldelimiterspace} \sigma _{\omega _{0} }^{2} } $. The $i$-th component is $d_{i} \in {\rm {\mathcal C}{\mathcal{N}}}\left(0,\sigma _{d_{i} }^{2} \right)$, $\sigma _{d_{i} }^{2} ={\rm {\mathbb{E}}}\left[\left|d_{i} \right|^{2} \right]$. \\ \hline 
${{\mathbf{y}}_{k}}$ & A $d$-dimensional zero-mean, circular symmetric complex Gaussian random vector, ${{\mathbf{y}}_{k}}\in \mathcal{C}\mathcal{N}\left( 0,\mathbb{E}\left[ {{\mathbf{y}}_{k}}\mathbf{y}_{k}^{\dagger } \right] \right)$. \\ \hline 
$y_{k,m} $ & The $m$-th element of the $k$-th user's vector ${{\mathbf{y}}_{k}}$, expressed as $y_{k,m} =\sum _{l}F\left(T_{i} \left({\rm {\mathcal N}}_{i} \right)\right) {\kern 1pt} {\kern 1pt} F\left(d_{i} \right)+F\left(\Delta _{i} \right).$ \\ \hline 
$F\left( \mathbf{T}\left( \mathcal{N} \right) \right)$ & Fourier transform of $\mathbf{T}\left( \mathcal{N} \right)={{\left[ {{T}_{1}}\left( {{\mathcal{N}}_{1}} \right)\ldots ,{{T}_{l}}\left( {{\mathcal{N}}_{l}} \right) \right]}^{T}}\in {{\mathcal{C}}^{l}}$, the complex transmittance vector. \\ \hline 
$F\left(\Delta \right)$ & Complex vector, expressed as $F\left(\Delta \right)=e^{{\textstyle\frac{-F\left(\Delta \right)^{T} K_{F\left(\Delta \right)} F\left(\Delta \right)}{2}} } ,$ with covariance matrix ${{\mathbf{K}}_{F\left( \Delta  \right)}}=\mathbb{E}\left[ F\left( \Delta  \right)F{{\left( \Delta  \right)}^{\dagger }} \right]$. \\ \hline 
$\mathbf{y}\left[ j \right]$ & AMQD block, $\mathbf{y}\left[ j \right]=F\left( \mathbf{T}\left( \mathcal{N} \right) \right)F\left( \mathbf{d} \right)\left[ j \right]+F\left( \Delta  \right)\left[ j \right]$. \\ \hline 
$\tau ={{\left\| F\left( \mathbf{d} \right)\left[ j \right] \right\|}^{2}}$ & An exponentially distributed variable, with density $f\left(\tau \right)=\left({1 \mathord{\left/{\vphantom{1 2\sigma _{\omega }^{2n} }}\right.\kern-\nulldelimiterspace} 2\sigma _{\omega }^{2n} } \right)e^{{-\tau  \mathord{\left/{\vphantom{-\tau  2\sigma _{\omega }^{2} }}\right.\kern-\nulldelimiterspace} 2\sigma _{\omega }^{2} } } ,$${\rm {\mathbb{E}}}\left[\tau \right]\le n2\sigma _{\omega }^{2} $. \\ \hline 
$\sigma _{\omega }^{2} $ & Average quadrature modulation variance of the Gaussian subcarriers, $\sigma _{\omega }^{2} ={\textstyle\frac{1}{n}} \sum _{i=1}^{n}\sigma _{\omega _{i} }^{2}  =\sigma _{\omega _{0} }^{2} .$ \\ \hline 
$T_{Eve,i} $ & Eve's transmittance on the Gaussian sub-channel ${\rm {\mathcal N}}_{i} $, $T_{Eve,i} =\text{Re}T_{Eve,i} +{\rm i}\text{Im}T_{Eve,i} \in {\rm {\mathcal C}}$, $0\le \text{Re}T_{Eve,i} \le {1 \mathord{\left/{\vphantom{1 \sqrt{2} }}\right.\kern-\nulldelimiterspace} \sqrt{2} } $, $0\le \text{Im}T_{Eve,i} \le {1 \mathord{\left/{\vphantom{1 \sqrt{2} }}\right.\kern-\nulldelimiterspace} \sqrt{2} } $, $0\le \left|T_{Eve,i} \right|^{2} <1$. \\ \hline 
$\mathbf{M}_{\mathcal{A}}^{\left[ j \right]}$ & Rate selection matrix for the $j$-th AMQD block. It allocates the subcarriers to set ${\rm {\mathcal A}}$ of transmit users. \\ \hline 
$H_{diff} \left(x\right)$ & Differential entropy of the continuous-variable $x$. \\ \hline 
$H_{diff} \left(\left. x\right|y\right)$ & Conditional differential entropy for input continuous-variable $x$, and output continuous-variable $y$.  \\ \hline 
$C_{sum} \left({\rm {\mathcal N}}\right)$ & Sum capacity, the total throughput over the $l$ sub-channels of ${\rm {\mathcal N}}$ at a constant modulation variance $\sigma _{\omega }^{2} $. \\ \hline 
$C_{sym} \left({\rm {\mathcal N}}\right)$ & Symmetric capacity, the maximum common rate at which all users can reliably transmit information over the $l$ sub-channels of ${\rm {\mathcal N}}$. \\ \hline 
$R_{k} $ & Transmission rate of user $U_{k} $. \\ \hline 
$R_{sum} \left({\rm {\mathcal N}}\right)$ & Sum rate, the total rate over the $l$ sub-channels of ${\rm {\mathcal N}}$ at a constant modulation variance $\sigma _{\omega }^{2} $. \\ \hline 
$R_{sym} \left({\rm {\mathcal N}}\right)$ & Symmetric rate, the common rate at which all users can reliably transmit information over the $l$ sub-channels of ${\rm {\mathcal N}}$. \\ \hline 
${\rm{C}}$ & The ${\rm {\mathcal H}}$ convex hull of independent input distributions. \\ \hline 
$C_{1} ,\ldots ,C_{K} $ & Corner points of the capacity region ${\rm{C}}$ of $K$ users, $U_{1,\ldots ,K} $. \\ \hline 
$d_{i} $ & A $d_{i} $ subcarrier in an AMQD block. For subset ${\rm {\mathcal A}}\subseteq K$ with $\left|{\rm {\mathcal A}}\right|$ users and $n$ Gaussian sub-channels for the transmission, $d_{i} ={\textstyle\frac{1}{\sqrt{n} }} \sum _{k=0}^{\left|{\rm {\mathcal A}}\right|-1}z_{k}  e^{\frac{-{\rm i}2\pi ik}{n} } ,i=0,\ldots ,n-1$. \\ \hline 
$\xi $ & The mean square estimation error of the input $d_{i} $ from output ${{d}_{i}}^{\prime } $ at modulation variance $2\sigma _{\omega }^{2} $. For an ideal Gaussian input $d_{i} $, $\xi \left(d_{i} \right)={\textstyle\frac{1}{1+2\sigma _{\omega }^{2} }} ,$$\xi ^{-1} \left(\mho \right)={\textstyle\frac{1-\mho }{\mho }} .$ \\ \hline 
$\nu _{\min } $ & The $\min \left\{\nu _{1} ,\ldots ,\nu _{l} \right\}$ minimum of the $\nu _{i} $ sub-channel coefficients, where $\nu _{i} ={\sigma _{{\rm {\mathcal N}}}^{2}  \mathord{\left/{\vphantom{\sigma _{{\rm {\mathcal N}}}^{2}  \left|F\left(T_{i} \left({\rm {\mathcal N}}_{i} \right)\right)\right|^{2} }}\right.\kern-\nulldelimiterspace} \left|F\left(T_{i} \left({\rm {\mathcal N}}_{i} \right)\right)\right|^{2} } $ and $\nu _{i} <\nu _{Eve} $. \\ \hline 
${\rm {\mathcal G}}\left(\delta \right)_{p\left(x\right)} $ & Quantifies the deviation of the input modulation (distribution) from the ideal Gaussian random distribution, in terms of mean square estimation error, expressed as ${\rm {\mathcal G}}\left(\delta \right)_{p\left(x\right)} ={\textstyle\frac{1}{\nu _{\min } \kappa }} -\xi ^{-1} \left(\delta \right)_{p\left(x\right)} $, $\kappa ={\textstyle\frac{1}{\nu _{Eve} +\nu _{\min } \left(1-{\rm {\mathcal G}}\left(\delta \right)_{p\left(x\right)} \right)}} $. \newline For an ideal Gaussian input modulation, ${\rm {\mathcal G}}^{{\rm *}} \left(\delta \right)_{p\left(x\right)} =1$ for arbitrary $\delta $, while for a $p\left(x\right)$ distribution $0\le \delta \le 1$, ${\rm {\mathcal G}}\left(\delta \right)_{p\left(x\right)} ={\textstyle\frac{1}{\nu _{\min } \kappa }} -\xi ^{-1} \left(\delta \right)_{p\left(x\right)} $.  \\ \hline 
$\sigma _{\omega }^{2} $ & Modulation variance, $\sigma _{\omega }^{2} =\nu _{Eve} -\nu _{\min } {\rm {\mathcal G}}\left(\delta \right)_{p\left(x\right)} $, where $\nu _{Eve} ={\textstyle\frac{1}{\lambda }} $, $\lambda =\left|F\left(T_{{\rm {\mathcal N}}}^{{\rm *}} \right)\right|^{2} ={\textstyle\frac{1}{n}} \sum _{i=0}^{n-1}\left|\sum _{k=0}^{n-1}T_{k}^{{\rm *}} e^{{\textstyle\frac{-{\rm i}2\pi ik}{n}} }  \right|^{2}  $ and $T_{{\rm {\mathcal N}}}^{{\rm *}} $ is the expected transmittance of the Gaussian sub-channels under an optimal Gaussian collective attack. \\ \hline 
$\xi ^{-1} \left(\delta \right)_{p\left(x\right)} $ & For an ideal Gaussian input ${{\xi }^{\text{*}}}^{-1}{{\left( \delta  \right)}_{p\left( x \right)}}=\tfrac{1-{{\nu }_{\min }}\kappa }{{{\nu }_{\min }}\kappa }$, and ${{\xi }^{-1}}{{\left( \delta  \right)}_{p\left( x \right)}}>{{\xi }^{\text{*}}}^{-1}{{\left( \delta  \right)}_{p\left( x \right)}}$ for an arbitrary $p\left( x \right)$ distribution. \\ \hline 
$\nu _{\kappa } $ & Additional sub-channel coefficient for the correction of modulation imperfections. For an ideal Gaussian modulation, $\nu _{\kappa } =0$, while for an arbitrary $p\left(x\right)$  distribution $\nu _{\kappa } =\nu _{\min } \left(1-{\rm {\mathcal G}}\left(\delta \right)_{p\left(x\right)} \right)$, where $\kappa ={\textstyle\frac{1}{\nu _{Eve} -\nu _{\min } \left({\rm {\mathcal G}}\left(\delta \right)_{p\left(x\right)} -1\right)}} $. \\ \hline 
$O_{k,i} =\sqrt{a_{k,i} } e^{{\rm i}\theta _{k,i} } $ & Complex variable, characterizes the opportunistic Gaussian modulation, $a_{k,i} \ge 0$, $\theta _{k,i} \in \left[0,2\pi \right]$, $\sum _{s}a_{k,i}  >1$ for each allowed $U_{k} $ from subset ${\rm {\mathcal A}}$.  \\ \hline 
${{\mathbf{O}}_{k}}\left[ j \right]$ & An $s$-dimensional vector of the opportunistic modulation of $U_{k} $, $\left[\sqrt{a_{k,1} } e^{{\rm i}\theta _{k,1} } ,\ldots ,\sqrt{a_{k,s} } e^{{\rm i}\theta _{k,s} } \right]^{T} $, for the $j$-th AMQD block. \\ \hline 
$c$ & Average of the stationary distributed $T_{k} \left({\rm {\mathcal N}}_{U_{k} } \right)$ transmittance coefficients, $c={\textstyle\frac{1}{K}} \sum _{k=1}^{K}{\textstyle\frac{1}{d}} \sum _{j=1}^{d}\left|F\left(T_{k} \left({\rm {\mathcal N}}_{U_{k} } \right)\left[j\right]\right)\right|  $. \\ \hline 
${{\mathbf{{z}'_{k}}}}\left[ j \right]$ & A $d$-dimensional input vector in opportunistic Gaussian modulation, where $s$ is the number of Gaussian subcarriers of $U_{k} $ in the $j$-th AMQD block, ${{\mathbf{{z}'_{k}}}}\left[ j \right]={{\mathbf{O}}_{k}}{{\mathbf{z}}_{k}}\left[ j \right]=\sqrt{{{a}_{k,i}}}{{e}^{i{{\theta }_{k,i}}}}{{z}_{k,i}},\text{ }i=1,\ldots s$. \\ \hline 
${\rm {\mathcal N}}_{U_{k} } \left[j\right]$ & The set of ${\rm {\mathcal N}}_{i} $ Gaussian sub-channels from the set of $l$ good sub-channels that transmit the $s$ subcarriers of user $U_{k} $ in the $j$-th AMQD block, ${\rm {\mathcal N}}_{U_{k} } \left[j\right]=\left[{\rm {\mathcal N}}_{1} ,\ldots ,{\rm {\mathcal N}}_{s} \right]^{T} $. \\ \hline 
\end{longtable}
\end{center}
\end{document}